%% file: main.tex
\documentclass[manuscript,screen,acmsmall]{acmart}
\AtBeginDocument{%
  }

\setcopyright{acmlicensed}
\copyrightyear{2026}
\acmYear{2026}
\acmDOI{XXXXXXX.XXXXXXX}
\acmISBN{.}

\usepackage[normalem]{ulem}
\useunder{\uline}{\ul}{}

\usepackage{hyperref}
\usepackage{pifont}
\usepackage{amsfonts, amsthm, array}
\usepackage[ruled,vlined,linesnumbered]{algorithm2e}
\usepackage{enumitem,multirow,graphicx,subcaption,multicol,lipsum,float,adjustbox}
\usepackage{cleveref}
\usepackage{algorithmicx}
\usepackage[page]{appendix} 
\crefformat{section}{\S#2#1#3}
\crefformat{subsection}{\S#2#1#3}
\crefformat{subsubsection}{\S#2#1#3}
\usepackage{arydshln}
\usepackage{epstopdf}
\usepackage{pgfplots}
\pgfplotsset{compat=1.18} 
\usepackage{subcaption}
\newcolumntype{P}{>{\raggedright\arraybackslash}m{0.98\linewidth}}
\newcolumntype{C}{>{\arraybackslash}m{1.\linewidth}}

\usepackage{makecell}
\usepackage{placeins}
\definecolor{softblue}{RGB}{185, 205, 230}  
\usepackage{xcolor}

\newif\ifrevision
\revisiontrue  
\ifrevision
    \newcommand{\rev}[1]{\textcolor{black}{#1}}
    \newcommand{\revcolor}{\color{black}}
\else
    \newcommand{\rev}[1]{#1}
    \newcommand{\revcolor}{}
\fi

\begin{document}


\title{Improving Scientific Document Retrieval with Academic Concept Index}

\author{Jeyun Lee}
\authornote{Both authors contributed equally to this research.}
\email{newejing77@korea.ac.kr}
\author{Junhyoung Lee}
\authornotemark[1]
\email{jun011017@korea.ac.kr}
\affiliation{%
    \institution{Korea University}
    \city{Seoul}
    \country{South Korea}
}

\author{Wonbin Kweon}
\affiliation{
    \institution{University of Illinois Urbana-Champaign}
    \city{Champaign}
    \country{United States}
}
\email{wonbin@illinois.edu}

\author{Bowen Jin}
\affiliation{
    \institution{University of Illinois Urbana-Champaign}
    \city{Champaign}
    \country{United States}
}
\email{bowenj4@illinois.edu}

\author{Yu Zhang}
\affiliation{
    \institution{Texas A\&M University}
    \city{College Station}
    \country{United States}
}
\email{yuzhang@tamu.edu}

\author{Susik Yoon}
\affiliation{%
    \institution{Korea University}
    \city{Seoul}
    \country{South Korea}}
\email{susik@korea.ac.kr}

\author{Dongha Lee}
\affiliation{
    \institution{Yonsei University}
        \city{Seoul}
    \country{South Korea}
}
\email{donalee@yonsei.ac.kr}

\author{Hwanjo Yu}
\affiliation{
    \institution{Pohang University of Science and Technology}
    \city{Pohang}
    \country{South Korea}
}
\email{hwanjoyu@postech.ac.kr}

\author{Jiawei Han}
\affiliation{
    \institution{University of Illinois Urbana-Champaign}
    \city{Champaign}
    \country{United States}
}
\email{hanj@illinois.edu}

\author{SeongKu Kang}
\affiliation{%
    \institution{Korea University}
    \city{Seoul}
    \country{South Korea}}
\authornote{Corresponding author}
\email{seongkukang@korea.ac.kr}

\renewcommand{\shortauthors}{Jeyun Lee and Junhyoung Lee et al.}

\begin{abstract}
\input{./sections/000abstract}
\end{abstract}

\input{dfn}

\maketitle

\section{Introduction}
\input{./sections/010introduction.tex}

\section{Related Work}
\label{sec:preliminary}
\input{./sections/020Preliminary}

\section{Academic Concept Index Construction}
\label{sec:method1}
\input{./sections/040Method_ConceptIndex}

\section{\proposed: Concept Coverage-based Query set Generation}
\label{sec:method2}
\input{./sections/041Method_CCQGen}

\section{\proposedtwo: Concept-focused Context Expansion for Retrieval}
\label{sec:method3}
\input{./sections/042Method_CCExpand}

\section{Experiments}
\subsection{Experimental Setup}
\label{sec:experimentsetup}
\input{./sections/050Experiment_setup}

\input{./sections/051Experiment_result}

\input{./sections/052Experiment_result2}


\section{Conclusion}
\label{sec:conclusion}
\input{sections/070conclusion}

\vspace{0.3cm}
\noindent
\textbf{Acknowledgements.}
This work was the result of project supported by KT(Korea Telecom)-Korea University AICT R\&D Center.
This work was also supported by ICT Creative Consilience Program through the IITP grant funded by the MSIT (IITP-2026-RS-2020-II201819), the NRF grant funded by the MSIT (RS-2026-25486220), and Basic Science Research Program through the NRF funded by the Ministry of Education (NRF-2021R1A6A1A03045425), and the IITP grant funded by the MSIT (IITP-2026-RS-2025-02304828).

\bibliographystyle{ACM-Reference-Format}
\bibliography{main}

\newpage

\appendix
\input{sections/080Appendix}

\end{document}

%% file: sections/000Abstract.tex
Adapting general-domain retrievers to scientific domains is challenging due to the scarcity of large-scale domain-specific relevance annotations and the substantial mismatch in vocabulary and information needs.
Recent approaches address these issues through two independent directions that leverage large language models (LLMs): (1) generating synthetic queries for fine-tuning, and (2) generating auxiliary contexts to support relevance matching.
However, both directions overlook the diverse academic concepts embedded within scientific documents, often producing redundant or conceptually narrow queries and contexts.
To address this limitation, we introduce an \textbf{academic concept index}, which extracts key concepts from papers and organizes them guided by an academic taxonomy.
This structured index serves as a foundation for improving both directions.
First, we enhance the synthetic query generation with concept coverage-based generation (\textit{\proposed}), which adaptively conditions LLMs on uncovered concepts to generate complementary queries with broader concept coverage.
Second, we strengthen the context augmentation with concept-focused auxiliary contexts (\textit{\proposedtwo}), which leverages a set of document snippets that serve as concise responses to the concept-aware \proposed queries.
Extensive experiments show that incorporating the academic concept index into both query generation and context augmentation leads to higher-quality queries, better conceptual alignment, and improved retrieval performance.

%% file: dfn.tex
\newcommand{\proposed}{CCQGen\xspace}
\newcommand{\proposedtwo}{CCExpand\xspace}

\newcommand{\smallsection}[1]{{\vspace{0.03in} \noindent \bf {#1.}}}

\newcommand{\ctr}{{Contriever-MS}\xspace}
\newcommand{\specter}{{SPECTER-v2}\xspace}
\newcommand{\csfcube}{CSFCube\xspace}
\newcommand{\dorismae}{DORIS-MAE\xspace}

%% file: sections/010introduction.tex
Scientific document retrieval is a fundamental task that supports scientific progress by enabling efficient access to technical knowledge~\cite{taxoindex, corank, kweon2026pairsem, semrank}.
Recently, pre-trained language model (PLM)-based retrievers~\cite{CTR, DPR} have shown strong performance in general-domain search tasks, benefiting from pre-training followed by fine-tuning on annotated query–document pairs.
However, adapting these general-domain retrievers to scientific domains faces two major challenges.
First, acquiring large-scale domain-specific relevance annotations is \rev{extremely costly and requires substantial expert effort}~\cite{li2023sailer, ToTER, inpars}.
Second, scientific corpora differ markedly from general-domain data in both vocabulary and information needs, which often leads to degraded retrieval performance~\rev{\cite{ToTER, taxoindex}}.
%

With the rapid advancement of large language models (LLMs)~\cite{GPT3, FLAN, Tzero, Llama}, recent studies have explored two major directions to improve \rev{document retrieval.} 
The first direction is \textbf{synthetic query generation}~\rev{\cite{dai2022promptagator, pairwise_qgen}}.
LLMs are prompted to generate training queries for each document with an instruction such as ``generate five relevant queries to the document'' \cite{synthetic_apple_VA, dai2022promptagator}.
These generated queries act as proxies for real user queries.
Advances in prompting schemes, including example-augmented prompting~\cite{inpars, dai2022promptagator} and pair-wise generation~\cite{pairwise_qgen}, have further improved the quality and diversity of synthetic queries.
Retrievers fine-tuned on these improved queries often achieve higher performance on specialized corpora.

The second direction focuses on \textbf{generating auxiliary contexts} to bridge the vocabulary and semantic gap between queries and documents~\cite{mackie2023generative, hyde, doc2query, doc2querymm, query2doc, mao2021generation}.
For example, \rev{Mackie et al.}~\cite{mackie2023generative} \rev{expand} the original query by prompting LLMs to produce additional semantic cues such as relevant keywords, entities, and summaries.
\rev{Gao et al.}~\cite{hyde} \rev{generate} hypothetical documents that serve as pseudo-answers to the query, providing richer interpretations to the user’s information need while aligning the expression styles with actual documents.
These generated contexts provide valuable signals that help the retriever capture relevance beyond surface-level text similarity.
A notable advantage of this line of work is that it is \textit{training-free}, requiring no retriever fine-tuning and thus allowing flexible deployment across different corpora.
This property is particularly appealing for scientific retrieval, where research labs and institutions often maintain their own local scientific corpora with unique distributions and terminologies.
In such environments, retraining a dedicated retriever for every individual corpus is impractical, making training-free augmentation methods~especially~valuable.


Despite their effectiveness, both lines of research share a fundamental limitation: they do not explicitly incorporate the academic concepts embedded within scientific documents.
Academic concepts refer to fundamental ideas, theories, and methodologies that constitute the core content of scientific texts~\cite{taxoindex, ccqgen}.
A scientific document typically discusses multiple such concepts, spanning fundamental ideas, theories, and methodologies, and domain-specific challenges.
However, existing LLM-based approaches do not model these conceptual structures.
In the synthetic query generation line, LLMs often focus on only a narrow subset of the document’s concepts, resulting in redundant queries with limited concept coverage~\cite{ccqgen}.
Likewise, in the context-augmentation line, without direct guidance, the generated contexts often fail to reflect the diverse concepts in the document, yielding only a narrow contextual signal and consequently limited improvements.

\begin{figure}[t]
    \centering
    \includegraphics[width=\linewidth]{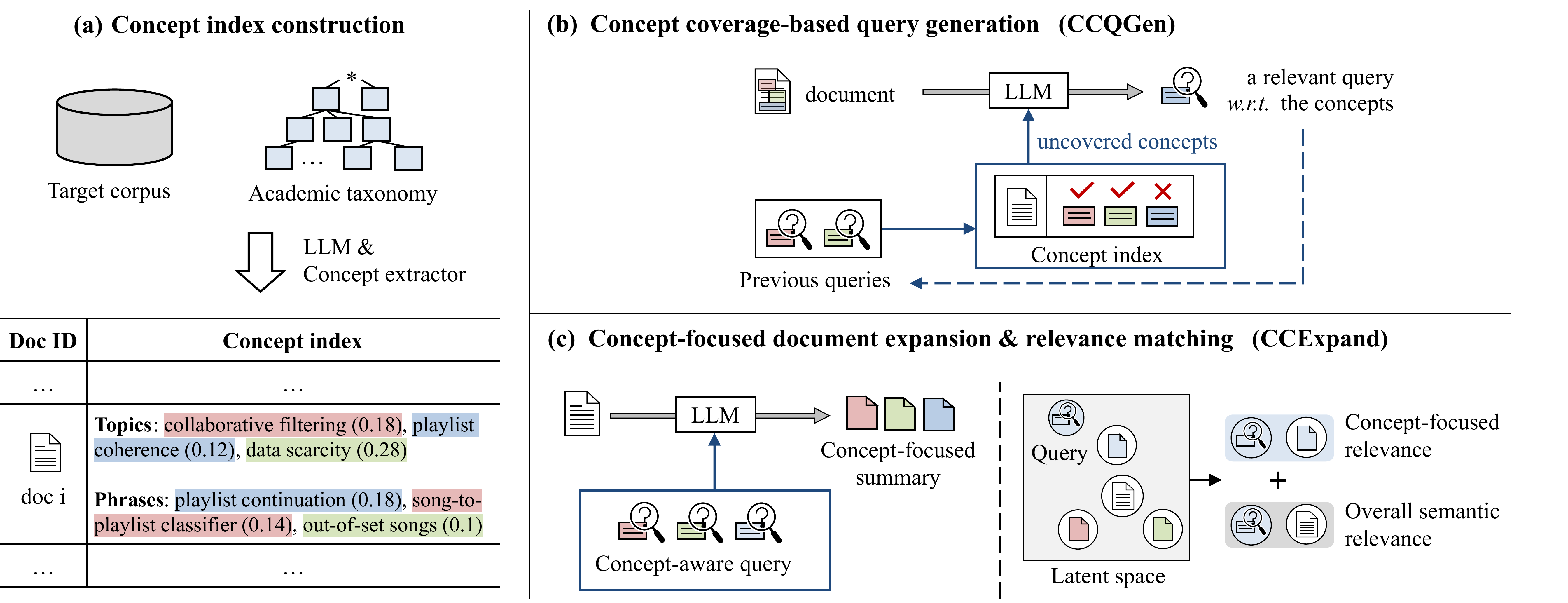}
    \caption{Overview of our framework. (a) We construct an academic concept index by extracting key concepts from each document.
    Leveraging this concept index, we explore two different directions for improving retrieval.
    (b) \proposed uses the index to guide synthetic query generation by identifying uncovered concepts and producing complementary queries.
    (c) \proposedtwo generates concept-focused snippets that act as concise, concept-grounded views of the document, enabling fine-grained concept matching.}
    \label{fig:intro}
\end{figure}

To address this limitation, we introduce an \textit{academic concept index} that provides a structured representation of the key concepts discussed within each document (Figure~\ref{fig:intro}a).
The index is constructed by extracting high-level topics, domain-specific phrases, and their related terminology using LLMs and a concept extractor.
This structured organization offers a principled view of the document’s conceptual space.
We aim to improve both query generation and context-augmentation directions based on this academic concept index.
To this end, we propose two methods that leverage the academic concept index: \textbf{\proposed} for concept-aware query generation, and \textbf{\proposedtwo} for concept-focused, training-free context augmentation.

The first method is \textit{CCQGen}, which enhances the synthetic query generation line by incorporating the academic concept index into the generation process~(Figure~\ref{fig:intro}b).
For each document, we identify the set of core concepts captured in the index and monitor which concepts are already represented in previously generated queries.
We then condition the LLM to generate additional queries targeting the remaining, uncovered concepts.
This adaptive procedure reduces redundancy, encourages broader concept coverage, and results in training queries that better reflect the diverse academic aspects of scientific documents.
By generating concept-aware and complementary queries, \proposed improves the effectiveness of fine-tuning retrievers on specialized scientific corpora.

The second method is \textit{\proposedtwo}, which strengthens context augmentation through concept-focused snippets extracted from the documents~(Figure~\ref{fig:intro}c).
Based on the concept index, \proposedtwo generates a set of targeted snippets that serve as concise responses to the concept-aware queries produced by \proposed.
These snippets provide \rev{concept-aware} views of the document, complementing the overall semantic similarity captured from surface text.
During retrieval, the query is matched not only against the global document representation but also against these concept-focused snippets, allowing the retriever to capture fine-grained conceptual signals without any additional model training.
Because \proposedtwo is fully training-free, it is easily deployable across diverse scientific corpora while providing structured, concept-grounded augmentation that enhances existing retrieval pipelines.

Our contributions are summarized as follows:
\begin{itemize}
    \item We introduce the academic concept index, a structured representation of key topics and domain-specific terminology extracted from each document.
    This index provides a foundation for concept-based retrieval enhancement.
    
    \item We propose \proposed, a concept-aware query generation method that adaptively conditions LLMs on uncovered concepts to produce complementary and diverse training queries.

    \item We propose \proposedtwo, a training-free, concept-focused context augmentation method that generates \rev{concept-aware} document snippets for fine-grained relevance matching.

    \item Through extensive experiments, we show that \proposed and \proposedtwo effectively improve retrieval performance, validating the importance of modeling the concept index in scientific document retrieval.
\end{itemize}

%% file: sections/020Preliminary.tex
PLM-based retrievers have become the standard foundation for modern retrieval, and many studies have sought to enhance them through two major directions: \textbf{synthetic query generation} and \textbf{auxiliary context augmentation}.
In this section, we review PLM-based retrieval and these two lines of work.

\subsection{PLM-based retrieval models}
The advancement of PLMs has led to significant progress in retrieval.
Recent studies have introduced retrieval-targeted pre-training \cite{CTR, condenser}, distillation from cross-encoders \cite{AR2}, and advanced negative mining methods \cite{zhan2021optimizing, rocketqa_v1, ContrastiveSurvey2023}.
There is also an increasing emphasis on pre-training methods specifically designed for the scientific domain. 
In addition to pre-training on academic corpora \cite{SCIBERT}, researchers have exploited metadata associated with scientific papers. 
\rev{Razdaibiedina and Brechalov}~\cite{razdaibiedina2023miread} use journal class, \rev{Cohan et al.}~\cite{SPECTER} \rev{and Ostendorff et al.}~\cite{SCINCL} use citations, \rev{Mysore et al.}~\cite{ASPIRE} use co-citation contexts, and \rev{Liu et al.}~\cite{OAGBERT} utilize venues, affiliations, and authors.
\rev{Singh et al.}~\cite{SPECTER2} \rev{and Zhang et al.}~\cite{zhang2023pre} devise multi-task learning of related tasks such as citation prediction~and~paper~classification.
Very recently, \rev{Kang et al.}~\cite{ToTER, taxoindex} leverage corpus-structured knowledge (e.g., core topics and phrases) for academic concept matching.
%

To perform retrieval on a new corpus, a PLM-based retriever is typically fine-tuned using a training set of annotated query-document pairs.
For effective fine-tuning, a substantial amount of training data is required. 
However, in specialized domains such as scientific document search, constructing vast human-annotated datasets is challenging due to the need for domain expertise, which remains an obstacle for~applications~\cite{li2023sailer, ToTER}.

\subsection{Synthetic query generation}
\label{prelim:qgen}
Earlier studies \cite{nogueira2019doc2query, liang2020embedding, ma2021zero, wang2022gpl} have employed dedicated query generation models, trained using massive document-query pairs from general domains.
Recently, there has been a shift towards replacing these generation models with LLMs \cite{dai2022promptagator, inpars, inpars2, pairwise_qgen, saad2023udapdr, sachan2022improving}.
Recent advancements have centered on improving prompting schemes to enhance the quality of these queries.
We summarize recent methods in terms of their prompting schemes.

\smallsection{Few-shot examples}
Several methods \cite{inpars, inpars2, dai2022promptagator, pairwise_qgen, label_condition_qgen, saad2023udapdr} incorporate a few examples of relevant query-document pairs in the prompt.
The prompt comprises the following components: $P = \{inst, (d_i, q_i)^k_{i=1}, d_t\}$, 
where $inst$ is the textual instruction\footnote{For example, ``\textit{Given a document, generate five search queries for which the document can be a perfect answer}''. 
The instructions vary slightly across methods, typically in terms of word choice.
In this work, we follow the instructions used in \cite{pairwise_qgen}.
}, 
$(d_i, q_i)^k_{i=1}$ denotes $k$ examples of the document and its relevant query, 
and $d_t$ is the new document we want to generate queries for.
By providing actual examples of the desired outputs, this technique effectively generates queries with distributions similar to actual queries (e.g., expression styles and lengths) \cite{dai2022promptagator}.
It is worth noting that this technique is also utilized in subsequent prompting schemes.

\smallsection{Label-conditioning} 
Relevance label $l$ (e.g., relevant and irrelevant) has been utilized to enhance query generation \cite{label_condition_qgen, saad2023udapdr, inpars}.
The prompt comprises $P = \{inst, (l_i, d_i, q_i)^k_{i=1}, (l_t, d_t)\}$, where $k$ label-document-query triplets are provided as examples.
$l_i$ represents the relevance label for the document $d_i$ and its associated query $q_i$.
To generate queries, the prompt takes the desired relevance label $l_t$ along with the document $d_t$.
This technique incorporates knowledge of different relevance, which aids in improving query quality and allows for generating both relevant and irrelevant queries \cite{label_condition_qgen}.

\smallsection{Pair-wise generation}
To further enhance the query quality, the state-of-the-art method \cite{pairwise_qgen} introduces a \textit{pair-wise} generation of relevant and irrelevant queries.
It instructs LLMs to first generate relevant queries and then generate relatively less relevant ones. 
The prompt comprises $P = \{inst, (d_i, q_i, q^-_i)^k_{i=1}, d_t\}$, where $q_i$ and $q^-_i$ denote relevant and irrelevant query for $d_i$, respectively.
The generation of irrelevant queries is conditioned on the previously generated relevant ones, allowing for generating thematically similar rather than completely unrelated queries.
These queries can serve as natural `hard negative' samples for training \cite{pairwise_qgen}.



While these prompting schemes enhance the realism and diversity of synthetic queries, they do not explicitly ensure coverage of the document’s academic concepts.
\rev{Subsequent methods have further explored adaptive query generation to improve efficiency and performance~\cite{tong2025igft, kim2025syntriever}, but they also overlook concept coverage at the document level.}
This limitation motivates our concept-aware query generation approach, \proposed, which leverages the academic concept index to generate more comprehensive query sets.


\subsection{Auxiliary context generation}
Another line of research enriches the retrieval process by generating auxiliary contexts for relevance matching.
A notable advantage of this line of work is that it is \textit{training-free}, requiring no retriever fine-tuning and allowing flexible deployment across different corpora.
Early studies explored methods such as pseudo-relevance feedback~\cite{DensePRF} and topic-based query enrichment~\cite{chen2018snippet}.
More recently, generative models, including LLMs, have been leveraged to create richer contextual signals.
GAR~\cite{mao2021generation} expands queries by generating in-domain contexts, such as answers, answer-containing sentences, or passage titles, and appends them to form generation-augmented queries.
GRF~\cite{mackie2023generative} proposes prompting LLMs to generate a diverse set of semantic signals, such as keywords, entities, pseudo-queries, and summaries, which are aggregated to refine the query representation.
This multi-signal augmentation allows the retriever to capture various semantic facets of the user intent.
Query2doc~\cite{query2doc} generates short document-like expansions conditioned on the query to enrich its semantic representation.
HyDE~\cite{hyde} also adopts a related strategy by generating hypothetical documents conditioned on the query.
These pseudo-documents resemble the style of scientific documents while introducing terms and expressions not explicitly present in the original query, thereby providing richer semantic evidence.
Beyond query-side augmentation, document-side expansion approaches have been widely explored.
Doc2query~\cite{doc2query} generates synthetic queries for each document and appends them to expand its lexical footprint.
Doc2query$--$~\cite{doc2querymm} improves this strategy by filtering low-quality generated queries to reduce noise in document expansion.
These approaches collectively demonstrate the utility of generated content for enhancing both query and document representations.

Such training-free approaches can be particularly appealing for scientific retrieval, where research institutions often maintain their own local scientific corpora with different distributions.
In these settings, retraining a dedicated retriever for every individual corpus is costly and impractical, making training-free augmentation methods especially valuable.
However, existing methods are not grounded in the conceptual structure of documents and therefore do not account for its diverse academic concepts, motivating concept-aware augmentation approaches such as our \proposedtwo.

%% file: sections/040Method_ConceptIndex.tex
\begin{figure*}[t]
\centering
\includegraphics[width=1.0\textwidth]{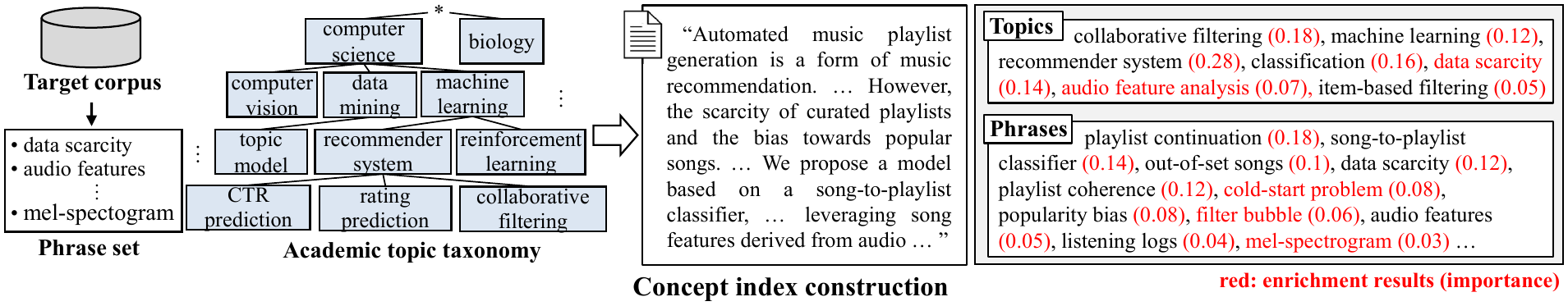}
\caption{Overview of the concept index construction. The illustration shows the indexing for a single document.}
\label{fig:method_index}
\end{figure*}

We construct an \textit{academic concept index} that captures the key concepts associated with each document (Figure \ref{fig:method_index}).
The index serves as the foundation for both directions explored in this work: concept-aware synthetic query generation (\proposed) and concept-focused context augmentation (\proposedtwo).
To build this index, we first identify the core academic concepts of each document at two levels of granularity (i.e., topics and phrases) and then enrich these concepts by estimating their importance and incorporating strongly related concepts that may not appear explicitly in~the~text.

\subsection{Concept Identification}
\label{subsec:method_a}
Scientific documents typically cover multiple academic concepts, ranging from broad research areas to fine-grained technical details.
To obtain a structured representation of these concepts, we identify core topics and core phrases for each document.
Topic-level concepts reflect high-level research themes such as ``collaborative filtering'' or ``machine learning'', while phrase-level concepts capture document-specific terminology such as ``playlist continuation'' or ``song-to-playlist classifier''.
These two levels provide complementary perspectives for characterizing the document’s content.

A straightforward approach is to instruct LLMs to list topics or phrases contained in the document.
However, such direct generation often exhibits two limitations:
the results may contain concepts not covered by the document, and there is always a potential risk of hallucination.
As a solution, we propose a new approach that first constructs a candidate set, and then uses LLMs to pinpoint the most relevant ones from the given candidates, instead of directly generating them.
By doing so, the output space is restricted to the predefined candidate space, greatly reducing the risk of hallucinations while effectively leveraging the language-understanding capability of LLMs.

\subsubsection{\textbf{Core topics identification}}
\label{subsub:core_topic}
To identify the core topics of documents, we propose using an \textit{academic topic taxonomy} \cite{MAG_FS}. 
In the scientific domain, academic taxonomies are widely used for categorizing studies in various institutions and can be easily obtained from the web.\footnote{E.g., IEEE Taxonomy (\href{https://www.ieee.org/content/dam/ieee-org/ieee/web/org/pubs/ieee-taxonomy.pdf}{\color{blue} link}), ACM Computing Classification System (\href{https://dl.acm.org/ccs}{\color{blue} link}).}
A taxonomy refers to a hierarchical tree structure outlining academic topics (Figure \ref{fig:method_index}).
Each node represents a topic, with child nodes corresponding to its sub-topics.
Leveraging the taxonomy allows for exploiting domain knowledge of topic hierarchy and reflecting researchers' tendency to~classify~studies.

\smallsection{Candidate set construction}
One challenge in finding candidate topics is that the taxonomy obtained from the web is often very large and contains many irrelevant topics.
To effectively narrow down the candidates, we employ a \textit{top-down traversal} technique that \textit{recursively visits} the child nodes with the highest \rev{similarities at each level (defined below)}.
For each document, we start from the root node and compute its similarity to each child node.
We then visit child nodes with the highest similarities.\footnote{We visit multiple child nodes and create multiple paths, as a document usually covers various topics. For a node at level $l$, we visit $l+2$ nodes to reflect the increasing number of nodes at deeper levels of the taxonomy. The root~node~is~level~$0$.}
This process recurs until every path reaches leaf nodes, and \textit{all visited nodes} are regarded as candidates.

The document-topic similarity ${s}(d, c)$ can be defined in various ways.
As a topic encompasses its subtopics, we collectively consider the subtopic information for each topic node.
Let $\mathcal{N}_c$ denote the set of nodes in the sub-tree having $c$ as a root node.
We compute the similarity as: ${s}(d, c) = \frac{1}{|\mathcal{N}_c|}\sum_{j \in \mathcal{N}_c} \operatorname{cos}(\mathbf{e}_{d}, \mathbf{e}_{j})$, 
where $\mathbf{e}_d$ and $\mathbf{e}_j$ denote representations from PLM for a document $d$ and the topic name of node $j$, respectively.\footnote{We use BERT with mean pooling as the simplest choice.}

\smallsection{Core topic selection}
We instruct LLMs to select topics most aligned with the document’s central theme from the candidate set.
An example of an input prompt is:
\begin{table}[h]
\small
    \centering
    \resizebox{1.0\linewidth}{!}{
    \begin{tabular}{|C|}
    \hline
    You will receive a document along with a set of candidate topics. Your task is to select the topics that best align with the core theme of the document. 
    Exclude topics that are too broad or less relevant.
    You may list up to [$k^t$] topics, using only the topic names in the candidate set. \textbf{Document}:~[\textsc{Document}],~\textbf{Candidate~topic~set}:~[\textsc{Candidates}]\\ \hline 
    \end{tabular}}
\end{table}

We set $k^t=10$. \rev{This value was determined empirically; performance remained stable across nearby values, indicating low sensitivity to this specific choice.}
For each document $d$, we obtain core topics as $\mathbf{y}^t_d \in \{0,1\}^{|\mathcal{T}|}$, where $y^t_{di}=1$ indicates $i$ is a core topic of $d$, otherwise $0$.
$\mathcal{T}$ denotes the topic set obtained~from~the~taxonomy.


%




%

\subsubsection{\textbf{Core phrases identification}}
\label{method:core_phrase}
From each document, we identify core phrases used to describe its concepts.
These phrases offer fine-grained details not captured at the topic level.
We note that not all phrases in the document are equally important.
Core phrases should describe concepts strongly relevant to the document but \textit{not frequently covered} by other documents with similar topics.
For example, among documents about `recommender system' topic, the phrase `user-item interaction' is very commonly used, and less likely to represent the most important concepts~of~the~document.

\smallsection{Candidate set construction}
Given the phrase set $\mathcal{P}$ of the corpus\rev{,}\footnote{The phrase set is obtained using an off-the-shelf phrase mining tool \cite{autophrase}.} we measure the distinctiveness of phrase $p$ in document $d$.
Inspired by recent phrase mining methods \cite{tao2016multi, lee2022taxocom}, we compute the distinctiveness as: $\exp(\operatorname{BM25}(p, d))/\,(1 + \sum_{d'\in\mathcal{D}_{d}}\exp(\operatorname{BM25}(p, d')))$.
This quantifies the relative relevance of $p$ to the document $d$ compared to other topically similar documents $\mathcal{D}_{d}$. 
$\mathcal{D}_{d}$ is simply retrieved using Jaccard similarity of core topics $\mathbf{y}^t_d$.
We set $|\mathcal{D}_{d}|=100$.
We select phrases with top-20\% distinctiveness score~as~candidates. 

\smallsection{Core phrase selection}
We instruct LLMs to select the most relevant phrases (up to $k^p$ phrases) from the candidates, using the same instruction format used for the topic selection.
We set $k^p=15$.
The core phrases are denoted by $\mathbf{y}^p_d \in \{0,1\}^{|\mathcal{P}|}$, where $y^p_{dj}=1$ indicates $j$ is a core phrase of $d$, otherwise $0$.

\subsection{\textbf{Concept Enrichment}}
\label{method:enrich}
We have identified core topics and phrases representing each document's concepts.
We further enrich this information by (1) measuring their relative importance, and (2) incorporating strongly related concepts (i.e., topics and phrases) not explicitly revealed in the document.
This enriched information serves as the basis for generating queries.

\vspace{0.02in} \noindent
\textbf{Concept extractor.}
We employ a small model called a \textit{concept extractor}.
For a document $d$, the model is trained to predict its core topics $\mathbf{y}^{t}_{d}$ and phrases $\mathbf{y}^{p}_{d}$ from the PLM representation $\mathbf{e}_d$.
We formulate this as a two-level classification task: topic and~phrase~levels.


Topics and phrases represent concepts at different levels of granularity, and learning one task can aid the other by providing a complementary perspective.
To exploit their complementarity, we employ a multi-task learning model with two heads \cite{mmoe}.
Each head has a Softmax output layer, producing probabilities for topics $\hat{\mathbf{y}}^{t}_{d}$ and phrases $\hat{\mathbf{y}}^{p}_{d}$, respectively.
The cross-entropy loss is then applied for classification learning: $-\sum_{i=1}^{|\mathcal{T}|} y^t_{di} \log \hat{y}^{t}_{di} - \sum_{j=1}^{|\mathcal{P}|} y^p_{dj} \log \hat{y}^{p}_{dj}$.

\smallsection{Concept enrichment}
Using the trained concept extractor, we compute $\hat{\mathbf{y}}^{t}_{d}$ and $\hat{\mathbf{y}}^{p}_{d}$, which reveal their importance in describing the document's concepts.
Also, we identify strongly related topics and phrases that are expressed differently or not explicitly mentioned, by incorporating those with the highest prediction probabilities.
For example, in Figure \ref{fig:method_index}, we identify phrases `cold-start problem', `filter bubble', and `mel-spectrogram', which are strongly relevant to the document's concepts but not explicitly mentioned, along with their detailed importance.
These phrases are used to aid in articulating the document's concepts in various related terms.


In sum, we obtain $k^{t'}$ enriched topics and $k^{p'}$ enriched phrases for each document with their importance from $\hat{\mathbf{y}}^{t}_{d}$ and $\hat{\mathbf{y}}^{p}_{d}$.
We set the probabilities for the remaining topics and phrases as $0$, and normalize the probabilities for selected topics and phrases, denoted by $\bar{\textbf{y}}^t_d$ and $\bar{\textbf{y}}^p_d$.
These enriched concept representations collectively form the \textbf{academic concept index}, which serves as the foundation for the two different directions explored in this work: \proposed and \proposedtwo.


%% file: sections/041Method_CCQGen.tex
We introduce \proposed, which enhances synthetic query generation by explicitly grounding the process in the academic concept index.
\proposed is designed to meet two desiderata for high-quality training queries:
(1) the generated queries should collectively provide complementary coverage of the document’s concepts, and
(2) each query should articulate the concepts using diverse and conceptually related terminology, rather than repeating surface-level phrases from the document.

Given previously generated queries $Q^{m-1}_d = \{q^1_d, ..., q^{m-1}_d\}$, we compare their concept coverage with the document’s concept index to identify concepts that remain insufficiently covered.
These uncovered concepts are then encoded as explicit conditions in the prompt to guide the LLM toward producing the next query $q^{m}_d$.
A final filtering stage ensures the quality and conceptual correctness of the generated queries.
This process is repeated until a predefined number ($M$) of queries per document is achieved.
$M$ is empirically determined, considering available training resources such as GPU memory and training time.
For the first query of each document, we impose no conditions, thus it is identical to the results obtained from existing methods.
Figure \ref{fig:method_ccqgen} illustrates the workflow.

\subsection{Identifying Uncovered Concepts}
\label{subsec:method_b}
The enriched concept index provides distributions $\bar{\mathbf{y}}^t_d$ and $\bar{\mathbf{y}}^p_d$ indicating the relative importance of topics and phrases for document $d$.
Our key idea is to generate queries that align with this distribution to ensure comprehensive coverage of the document's concepts.
To estimate the concepts already expressed in earlier queries, we apply the concept extractor to the concatenated text $Q^{m-1}_d$, yielding $\bar{\mathbf{y}}^t_Q$ and $\bar{\mathbf{y}}^p_Q$.

Based on the concept coverage information, we identify concepts that need to be more emphasized in the subsequently generated query.
A concept is ``under-covered’’ if its value is high in $\bar{\mathbf{y}}^p_d$ but low in $\bar{\mathbf{y}}^p_Q$.
We formalize the under-coverage concept distribution as:
\begin{equation}
    \boldsymbol{\pi} = \operatorname{normalize}(\,\max(\bar{\textbf{y}}^p_d - \bar{\textbf{y}}^p_Q, \,\epsilon)\,)
\end{equation}
We set $\epsilon = 10^{-3}$ as a minimal value to the core phrases for numerical stability.
Here, we use phrase-level concepts for \textit{explicit} conditioning because they are more naturally incorporated into prompts than broad topic labels.
Note that topics are \textit{implicitly} reflected in identifying and enriching core phrases. 
We sample $\lfloor \frac{k^{p'}}{M} \rfloor$ phrases according to the multinomial distribution:
\begin{equation}
\mathcal{S} = \operatorname{Multinomial} \left(\left\lfloor \frac{k^{p'}}{M} \right\rfloor,  \boldsymbol{\pi}\right),
\end{equation}
where $\mathcal{S}$ denotes the set of sampled phrases and $M$ is the total number of queries to be generated for the document.
As queries accumulate, $\bar{\mathbf{y}}^p_Q$ updates dynamically, enabling progressive concept coverage.

\begin{figure*}[t]
\centering
\includegraphics[width=1.0\textwidth]{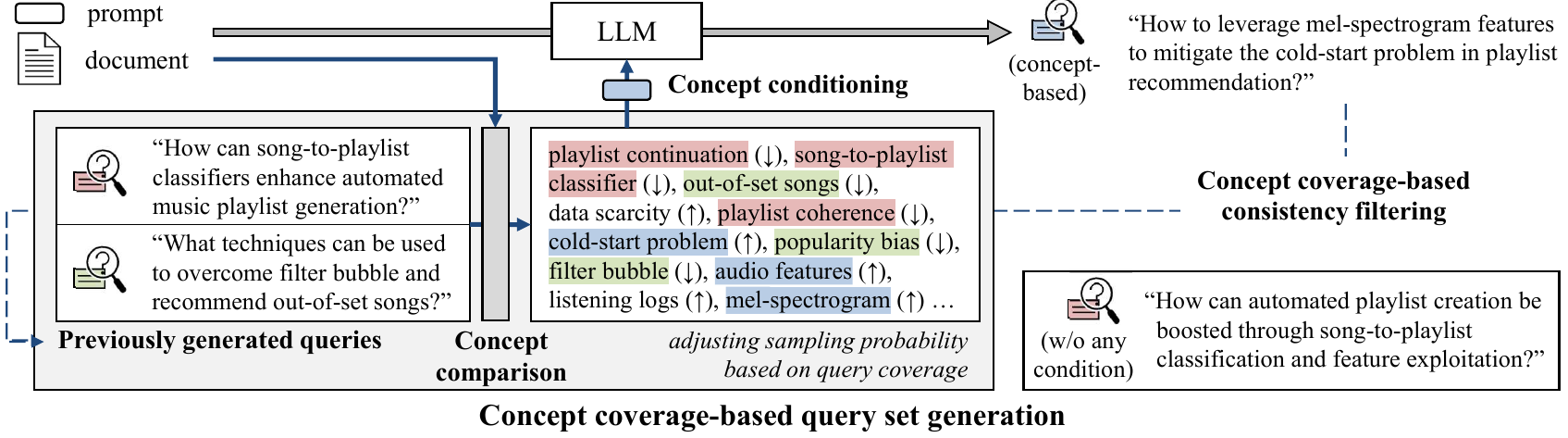}
\caption{The overview of Concept Coverage-based Query set Generation (\proposed).
Best viewed in color.
}
\label{fig:method_ccqgen}
\end{figure*}

To illustrate this process, consider the example in Figure~\ref{fig:method_ccqgen}.
Given the previously generated queries, the concept extractor identifies which concepts have already been expressed and which remain insufficiently covered.
For instance, phrases such as `playlist continuation', `song-to-playlist classifier,' and `out-of-set songs' exhibit relatively low under-coverage values (indicated by $\downarrow$), as they already appear in earlier queries.
In contrast, concepts such as `cold-start problem', `audio features', `mel-spectrogram', and `data scarcity' receive higher under-coverage values (indicated by $\uparrow$), reflecting their absence from prior queries despite their importance in the document.
Based on the distribution $\boldsymbol{\pi}$, \proposed increases the sampling probability of these uncovered concepts, making them more likely to be selected as conditioning phrases.

\subsection{\textbf{Concept-Conditioned Query Generation}}
\label{subsub:method_condition}
The sampled phrases are leveraged as conditions for generating the next query $q^m_d$.
Controlling LLM outputs through prompt-based conditioning has been widely studied in tasks such as sentiment control, keyword guidance, and outline-based text generation~\cite{control_gen, outline_condition}.
Following these studies, we impose a condition by adding a simple textual instruction $C$: ``\textit{Generate a relevant query based on the following keywords}: [\textsc{Sampled phrases}]''.
While more sophisticated instruction could be employed, we obtained satisfactory results with our choice.

Let $P$ denote any existing prompting scheme (e.g., few-shot examples) discussed in Section~\ref{prelim:qgen}.
The final prompt is constructed as $[P; C]$, and the LLM generates the next query according to:
\begin{equation}
q^m_d = \operatorname{LLM}([P; C]).
\label{eq:ccqgen_generation}
\end{equation}
This integration allows us to inherit the strengths of existing techniques, while explicitly steering the LLM toward under-covered concepts.
For example, in Figure \ref{fig:method_ccqgen}, the concept condition includes phrases like `cold-start problem' and `audio features', which are not well covered by the previous queries.
Based on this concept condition, we guide LLMs to generate a query that covers complementary aspects to the previous ones.
It is important to note that $C$ adds an \textit{additional condition} for $P$; the query is still about playlist recommendation, the main task of the document.

By contrast, when no concept condition is applied—as illustrated in the bottom-right of the figure—the LLM tends to produce queries that are fluent yet semantically redundant, often revisiting concepts already well covered by earlier queries.
This highlights the necessity of concept-conditioned prompting for systematically broadening conceptual coverage instead of relying solely on unconstrained LLM generation.

\smallsection{Concept coverage-based consistency filtering}
After generating a query, we apply a filtering step to ensure its quality.
A key criterion is \textit{round-trip consistency}~\cite{alberti2019synthetic}: the document that produced a query should also be retrieved by that query.
Prior work~\cite{dai2022promptagator, label_condition_qgen} implements this by retaining a query $q_d$ only if its source document $d$ appears within the top-$N$ results of a retriever.
However, this approach is often dominated by surface-level lexical matching.
In scientific domains, conceptually aligned queries frequently use terminology that differs from the exact phrasing in the document, causing retrievers to overlook semantically valid queries.

To address this issue, we incorporate concept-level similarity into the filtering stage.
For a query–document pair $(q_d,d)$, we compute textual similarity $s_{\text{text}}(q_d,d)$ from the retriever and concept similarity $s_{concept}(q_d,d) = sim(\bar{\mathbf{y}}^p_{q_d}, \,\bar{\mathbf{y}}^p_d)$, where $sim(\cdot)$ is inner-product similarity over the top 10\% phrases.
We then define a combined score for filtering:
\begin{equation}
\label{eq:dual_score}
s_{\text{filtering}}(q_d,d) = f(s_{\text{text}}(q_d,d),\ s_{\text{concept}}(q_d,d)),
\end{equation}
where $f$ denotes the function to consolidate two scores.
Here, we use simple addition after z-score normalization.
Round-trip consistency is assessed using this combined score.
By jointly considering surface-level and concept-level signals, this filtering more reliably preserves semantically meaningful queries that would otherwise be discarded due to lexical mismatch.
The resulting query–document pairs are used to fine-tune the retriever using a standard contrastive learning~\cite{DPR}.
\rev{Note that this filtering incurs minimal additional overhead, as the concept similarity scores are derived from the already trained concept extractor and do not require dedicated training or additional model components.}

\smallsection{Remarks on the efficiency of \proposed}
\proposed iteratively assesses the generated queries and imposes conditions for subsequent generations, which \rev{introduces additional computation during the offline query generation phase.
However, this overhead is limited to the query generation stage and does not introduce additional cost during fine-tuning or retrieval inference.
Further, \proposed yields more consistent improvements}, even when the number of queries is highly limited, whereas existing methods often fail to improve~the~retriever~(Section \ref{result:CCQGen}).

%% file: sections/042Method_CCExpand.tex
We now propose \proposedtwo, a training-free context augmentation method grounded in the academic concept index. 
While \proposed utilizes the concept index to generate training signals, \proposedtwo uses it to construct a set of document snippets, each focusing on complementary academic concepts covered in the document.
During retrieval, the query is matched not only against the document representation but also against these concept-focused snippets, enabling the retriever to capture fine-grained conceptual signals without any additional model training.


\begin{figure*}[t]
\centering
\begin{subfigure}[b]{1.0\textwidth}
    \centering
    \includegraphics[width=1.0\textwidth]{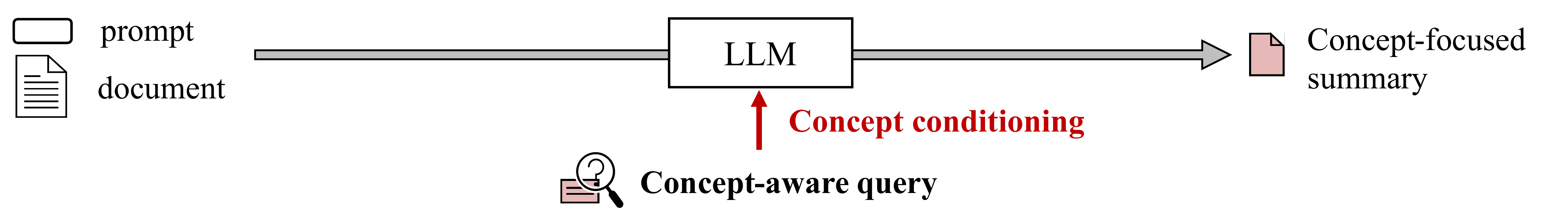}
    \caption{\rev{Overall pipeline of concept-focused snippet generation.}}
    \label{fig:ccexpand_gen_pipeline}
\end{subfigure}
\\[0.5em]
\begin{subfigure}[b]{1.0\textwidth}
    \centering
    \includegraphics[width=1.0\textwidth]{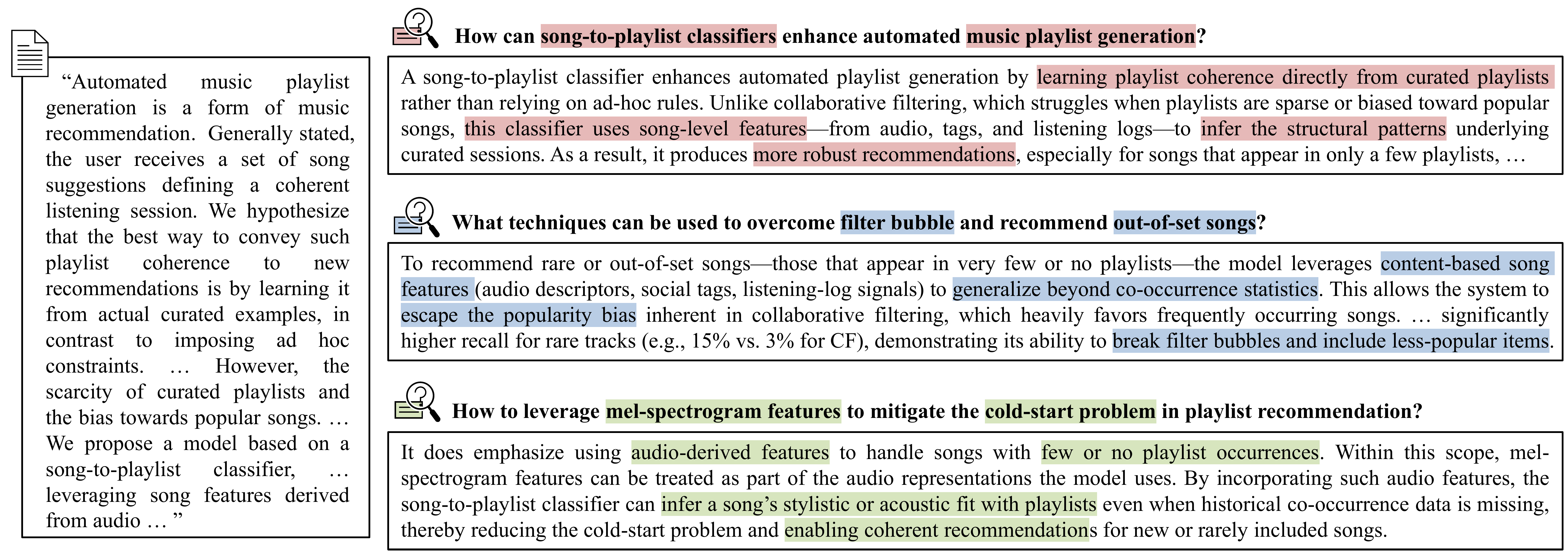}
    \caption{\rev{An illustrative example of generated concept-focused snippets.}}
    \label{fig:ccexpand_gen_example}
\end{subfigure}
\caption{An illustration of concept-focused snippet generation in \proposedtwo.
\rev{(a) shows the general pipeline, and (b) presents three snippet examples.}
Best viewed in color.
}
\label{fig:ccexpand_gen}
\end{figure*}

\subsection{Concept-focused Snippet Generation}
\label{sec:summary_gen}
A fundamental limitation of a dense retriever is that it compresses an entire document into a single vector representation, often resulting in the loss of fine-grained information~\cite{ASPIRE, taxoindex}.
This issue is particularly critical in scientific retrieval, where a single paper typically encompasses various academic concepts.
When these diverse concepts are blended into a single vector, the retriever struggles to distinguish \rev{between} the document’s conceptually distinct aspects.

To address this limitation, we leverage the academic concept index to construct a set of snippets that each reflect a different conceptual aspect of the document.
We note that, through \proposed, we can already generate a concept-aware query set that comprehensively covers the document’s key concepts.
Naturally, each query in $Q_d$ serves as a natural language instantiation of a specific subset of the concepts encoded in the document $d$.
Conditioned on each concept-aware query, we employ an LLM to generate a concept-focused snippet that tailors the document to answer that query.

Formally, the snippet $s_d$ corresponding to a query $q_d$ is generated as:
\begin{equation}
    s_d = \text{LLM}(inst, d, q_d),
\end{equation}
where $inst$ denotes the instruction. 
An example of an input prompt is:
\begin{table}[h]
\small
    \centering
    \resizebox{1.0\linewidth}{!}{
    \begin{tabular}{|C|}
    \hline
    Given a document and a query, generate one coherent paragraph (4-6 sentences) that explains how the document addresses the information need of the query, with emphasis on the concepts represented by the query's topics.
    \textbf{Document}:~[\textsc{Document}],~\textbf{Query}:~[\textsc{Query}]\\ \hline 
    \end{tabular}}
\end{table}

By repeating this process, we obtain a set of concept-focused snippets $S_d = \{s^1_d, \dots, s^M_d\}$, where $M$ is the number of generated queries per document.
Figure~\ref{fig:ccexpand_gen} illustrates this process.
Unlike generic augmentation methods~\cite{hyde, mackie2023generative}, which produce auxiliary content without explicit conceptual grounding, our approach leverages the concept index to ensure that the generated snippets collectively and systematically cover the diverse concepts present in the document.

\subsection{Concept-focused Relevance Matching}
\label{sec:matching}
As discussed earlier, compressing an entire scientific document into a single vector often fails to preserve its \rev{concept-aware} information, making it difficult for the retriever to answer queries targeting particular academic concepts.
We introduce a concept-focused relevance matching strategy that leverages the most aligned concept snippet as auxiliary context during retrieval.

\begin{figure*}[t]
\centering
\includegraphics[width=0.8\textwidth]{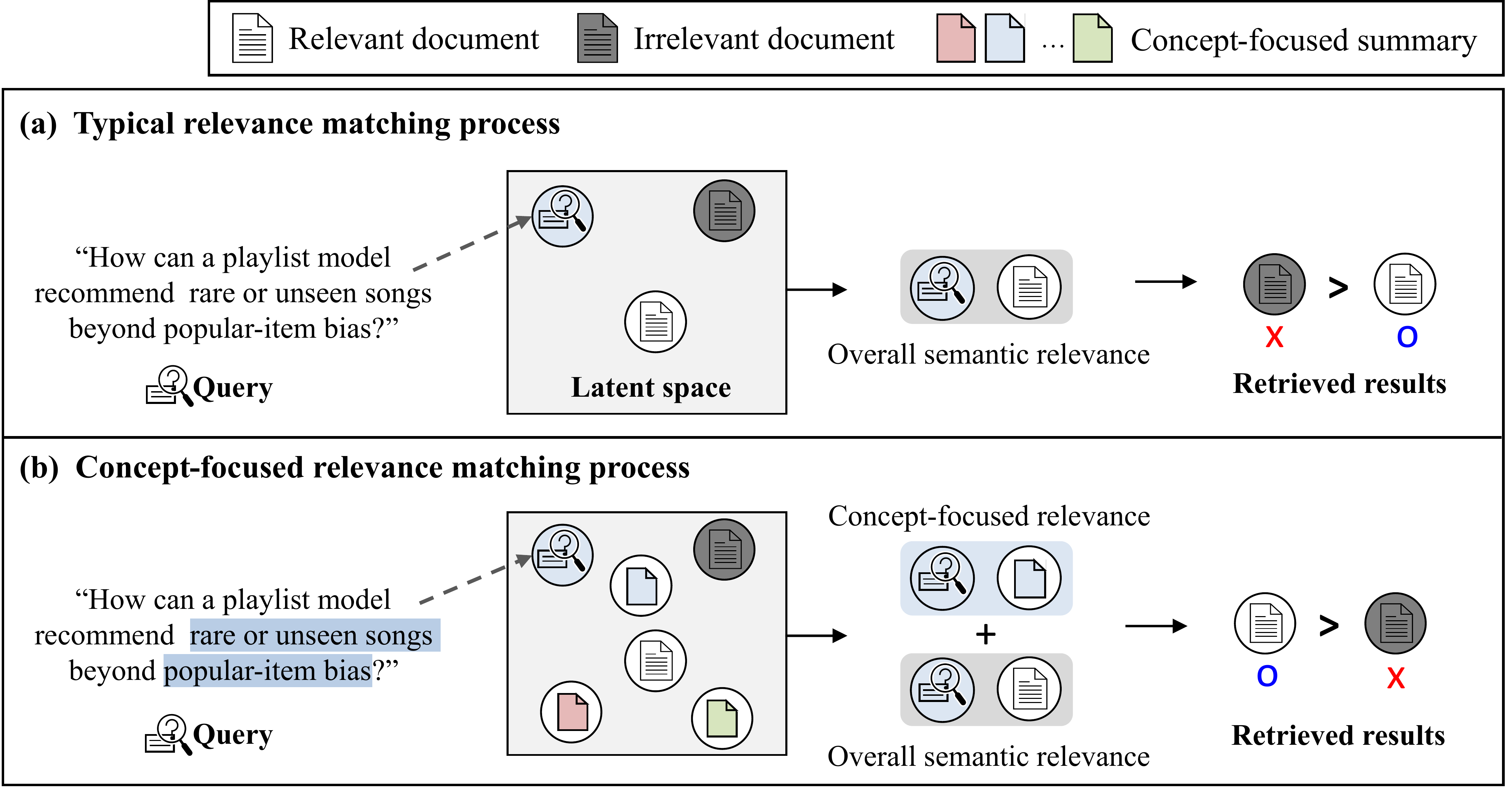}
\caption{An illustration of concept-focused relevance matching process.
Best viewed in color.
}
\label{fig:ccexpand_matching}
\end{figure*}

Given a query–document pair $(q_{test}, d)$, we first identify the snippet that best reflects the concepts emphasized in the query:
\begin{equation}
    s^*_{d} = \arg\max_{s \in S_d} (sim(q_{test}, s)),
\end{equation}
where $sim(\cdot,\cdot)$ denotes the retriever’s similarity score.
We then compute the refined relevance score by combining the overall textual similarity with the concept-focused similarity:
\begin{equation}
\text{rel}(q_{test}, d)
= f\Big( sim(q_{test}, d),\; sim(q_{test}, s_d^{*}) \Big)
\end{equation}
In practice, we use a simple linear combination, i.e., $f(a, b) = (1-\alpha)\, a + \alpha\, b$, where $\alpha \in [0, 1]$ is a hyperparameter that balances two views.
This process is illustrated in Figure~\ref{fig:ccexpand_matching}.

The proposed concept-focused matching offers several benefits.
First, it strengthens the relevance matching by injecting \rev{concept-aware} information that is otherwise diluted in a single document vector.
By matching the query against the snippet most aligned with its conceptual focus, the retriever can better distinguish documents that share similar global semantics but address different academic concepts.
Second, this approach is entirely training-free: it requires no modification or fine-tuning of the retriever, making it easily applicable across different scientific corpora and compatible with any dense retriever.

\smallsection{Discussion on inference latency}
A key advantage of \proposedtwo is that all snippet generation is performed entirely \emph{offline}.
During inference, the snippets are already vectorized and indexed for ANN search, requiring no LLM calls.
This contrasts sharply with recent online LLM-based augmentation methods~\cite{mackie2023generative, hyde}, whose latency is dominated by real-time prompting and decoding.
Moreover, concept-focused matching is applied only to a small subset of candidate documents rather than the full corpus.
Given a query, the retriever first computes standard similarity scores using the global document representations and selects the top-$K$ candidates (we use $K=1000$).
Concept-focused similarities are then computed only between the query and the snippets of these candidates.
Detailed efficiency analyses are provided in Section~\ref{subsubsec:CCExpand_result2}.


%% file: sections/050Experiment_setup.tex
\subsubsection{\textbf{Datasets}}
We conduct a thorough review of the literature to find retrieval datasets in the scientific domain, specifically those where relevance has been assessed by skilled experts or annotators.
We select two recently published datasets: \textbf{\csfcube} \cite{CSFCube} and \textbf{\dorismae} \cite{DORISMAE}.
They offer test query collections annotated by human experts and LLMs, respectively, and embody two real-world search scenarios: query-by-example and human-written queries.
For both datasets, we conduct retrieval from the entire corpus, including all candidate documents.
\csfcube dataset consists of 50 test queries, with about 120 candidates per query drawn from approximately 800,000 papers in the S2ORC corpus \cite{lo2020s2orc}. 
\dorismae dataset consists of 165,144 test queries, with candidates drawn similarly to \csfcube.
We consider annotation scores above `2', which indicate documents are `nearly identical or similar' (\csfcube) and `directly answer all key components' (\dorismae), as relevant.
Note that training queries are not provided in \rev{either} dataset.
\rev{Detailed data statistics are provided in Appendix~\ref{appendix:data_statistics}.}

\subsubsection{\textbf{Academic topic taxonomy}}
We utilize the field of study taxonomy from Microsoft Academic \cite{MAG_FS}, which contains $431,416$ nodes with a maximum depth of $4$.
After the concept identification step (Section \ref{subsec:method_a}), we obtain $1,164$ topics and $18,440$ phrases for \csfcube, and $1,498$ topics and $34,311$ phrases for \dorismae.

\subsubsection{\textbf{Metrics}}
Following \cite{mackie2023generative, ToTER}, we employ Recall@$K$ (R@$K$) for a large retrieval size ($K$), and NDCG@$K$ (N@$K$) and MAP@$K$ (M@$K$) for a smaller $K$ ($\leq 20$).
Recall@$K$ measures the proportion of relevant documents in the top $K$ results, while NDCG@$K$ and MAP@$K$ assign higher weights to relevant documents at higher~ranks.

\subsubsection{\textbf{Backbone retrievers}}
We employ two representative models: 
(1) \textbf{\ctr} \cite{CTR} is a widely used retriever fine-tuned using vast labeled data from general domains (i.e., MS MARCO).
(2) \textbf{\specter} \cite{SPECTER2} is a PLM specifically developed for the scientific domain. It is trained using metadata (e.g., citation relations) of scientific papers. 
For both models, we use public checkpoints: \texttt{facebook/contriever-msmarco} and  \texttt{allenai/specter2\_base}.

\subsubsection{\textbf{Implementation details}}
All experiments are conducted using PyTorch with CUDA.
For the concept extractor, we employ a multi-gate mixture of \rev{experts} architecture \cite{mmoe}, designed for multi-task learning.
We use three experts, each being a two-layer MLP.
We set the number of enriched topics and phrases to $k^{t'}=15$ and $k^{p'}=20$,~respectively. \rev{These values were chosen empirically through preliminary validation; we found that performance remained relatively stable across a reasonable range of values, indicating that the method is not overly sensitive to these specific choices.}

\smallsection{\proposed}
We conduct all experiments using 4 NVIDIA RTX A5000 GPUs, 512 GB memory, and a single Intel Xeon Gold 6226R processor. 
For fine-tuning, we use top-50 BM25 hard negatives for each query \cite{formal2022distillation}.
We use 10\% of training data as a validation set. 
The learning rate is set to $10^{-6}$ for \ctr and $10^{-7}$ for \specter, after searching among $\{10^{-7}, 10^{-6}, ..., 10^{-3}\}$.
We set the batch size as $64$ and the weight decay as $10^{-4}$.
We report the average performance over five independent runs.
For all query generation methods, we use \texttt{gpt-3.5-turbo-0125}.
For all methods, we generate five queries for each document ($M=5$).
For the few-shot examples in the prompt, we randomly select five annotated examples, which are then excluded in the evaluation process \cite{dai2022promptagator}.
We follow the textual instruction used in \cite{pairwise_qgen}.
For other baseline-specific setups, we adhere to the configurations described in the original papers.
For the consistency filtering, we set $N=5$.

\smallsection{\proposedtwo}
For all LLM-based baselines and \proposedtwo, we use \texttt{gpt-4o-mini}.\footnote{\texttt{gpt-3.5-turbo-0125} was categorized as a legacy model by OpenAI at the time of this study.}
In our experiments, we set $\alpha = 0.6$ for \csfcube and $\alpha = 0.4$ for \dorismae to balance the contributions of overall and concept-focused similarities.
We generate five snippets per document, matching the number of concept-aware queries produced by \proposed. 
For fair comparison across all context generation baselines, we keep both the number and type of auxiliary contexts consistent. Specifically, all document-expansion methods use five pseudo-queries, GRF~\cite{mackie2023generative} generates topics, keywords, and summaries, and HyDE~\cite{hyde} produces five hypothetical documents.

%% file: sections/051Experiment_result.tex
\input{sections/99991main_table}

\subsection{Experiment results for \proposed}
In this section, we present comprehensive experiments evaluating the effectiveness of \proposed.

\label{sec:experimentresult}
\subsubsection{\textbf{Compared methods}}
We compare various synthetic query generation methods. For each document, we generate \textbf{five} relevant queries~\cite{BEIR}.
\begin{itemize}[leftmargin=*]
    \item \textbf{GenQ} \cite{BEIR} employs a specialized query generation model, trained with massive document-query pairs from the general domains.
    We use T5-base, trained using approximately $500,000$ pairs from MS MARCO dataset \cite{nogueira2019doc2query}: \texttt{BeIR/query-gen-msmarco-t5-base-v1}.
\end{itemize}
\noindent
\proposed can be integrated with existing LLM-based methods to enhance the concept coverage of the generated queries.
We apply \proposed to two recent approaches, discussed~in~Section~\ref{prelim:qgen}.
\rev{These methods represent two distinct prompting strategies: few-shot example-based prompting and pair-wise generation. This selection is intended to validate the general applicability of CCQGen across diverse prompting schemes.
}
\begin{itemize}[leftmargin=*]
    \item \textbf{Promptgator} \cite{dai2022promptagator} is a recent LLM-based query generation method that leverages \textbf{few-shot examples} within the prompt. 
    

    \item \textbf{Pair-wise generation} \cite{pairwise_qgen} is the state-of-the-art method that generates relevant and irrelevant queries in a \textbf{pair-wise}~manner. \rev{For fine-tuning, both the positive (relevant) and negative (irrelevant) generated queries are incorporated as training data.}
\end{itemize}
Additionally, we devise a new competitor that adds more instruction in the prompt to enhance the quality of queries:
\textbf{Promptgator\_{diverse}} is a variant of Promptgator, where we add the instruction ``\textit{use various terms and reduce redundancy among the~queries}''.

\subsubsection{\textbf{Effectiveness of \proposed}}
\label{result:CCQGen}
Table \ref{tab:main} presents retrieval performance after fine-tuning with various query generation methods.
\proposed consistently outperforms all baselines, achieving significant improvements across various metrics with both backbone models.
We observe that GenQ underperforms compared to LLM-based methods, showing the advantages of leveraging the text generation capability of LLMs.
Also, existing methods often fail to improve the backbone model (i.e., no Fine-Tune), particularly \ctr.
As it is trained on labeled data from general domains, it already captures overall textual similarities well, making further improvements challenging.
The consistent improvements by \proposed support its efficacy in generating queries that effectively represent the scientific documents.
Notably, Promptgator\_diverse struggles to produce consistent improvements.
It often generates redundant queries covering similar aspects, despite increased diversity in their expressions (further analysis provided in Section \ref{result:query_analysis}).
This underscores the importance of proper control over generated content and supports the validity of our~approach.

\begin{figure}[t]
\centering
\includegraphics[height=3.2cm]{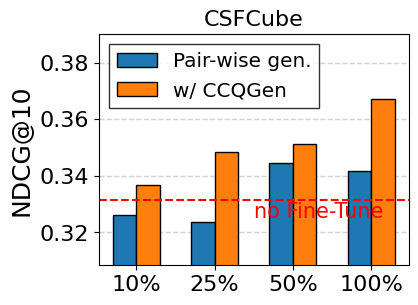}
\includegraphics[height=3.2cm]{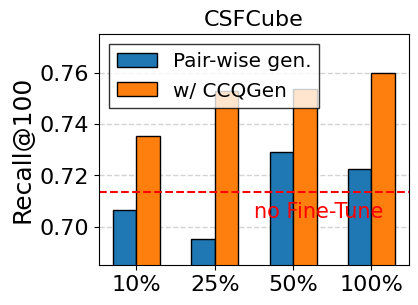} \\
\includegraphics[height=3.2cm]{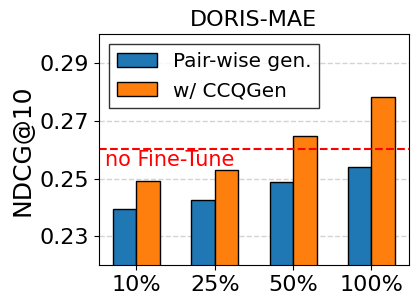}
\includegraphics[height=3.2cm]{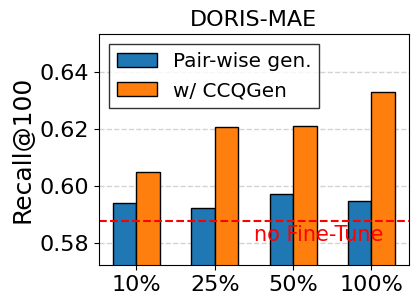} 
\caption{Results with varying amounts of training data.
x\% denotes setups using a random x\% of queries.
}
\label{fig:amount}
\end{figure}

\smallsection{\textbf{Impact of amount of training data}}
In Figure \ref{fig:amount}, we further explore the retrieval performance by limiting the amount of training data, using \ctr as the backbone model.
The existing LLM-based generation method (i.e., Pair-wise gen.) shows limited performance under restricted data conditions and fails to fully benefit from an increasing volume of training data.
This supports our claim that the generated queries are often redundant and do not effectively introduce new training signals.
Conversely, \proposed delivers \rev{relatively consistent and} considerable improvements, even with a limited number of queries.
\proposed guides each new query to complement the previous ones, allowing for reducing redundancy and fully leveraging the limited number of queries.

\begin{table*}[t]
\centering
\caption{Compatibility with retrieval enhancement methods.
Relative improvements (\%) over the retriever fine-tuned with \proposed. 
* indicates a significant difference (paired t-test, $p$ < 0.05) from the best baseline (i.e., Pair-wise generation with ToTER).}
\centering
\renewcommand{\arraystretch}{1}
\resizebox{0.85\linewidth}{!}{
\begin{tabular}{lcccccccc}
\toprule
& \multicolumn{4}{c}{\textbf{CSFCube}} & \multicolumn{4}{c}{\textbf{DORIS-MAE}} \\
\cmidrule(lr){2-5} \cmidrule(lr){6-9}
\textbf{Method} 
& N@10 & N@20 & M@10 & M@20 
& N@10 & N@20 & M@10 & M@20 \\
\midrule
CCQGen + GRF 
& +1.93\% & +0.20\% & +3.56\% & +1.97\%
& \textcolor{red}{-2.66\%} & \textcolor{red}{-0.61\%} & \textcolor{red}{-3.52\%} & \textcolor{red}{-2.22\%} \\

CCQGen + ToTER 
& +9.62\% & +3.49\% & +11.35\% & +7.85\%
& +1.28\% & +6.54\% & +7.34\% & +6.57\% \\

CCQGen++ 
& +\textbf{15.64}\%* & +\textbf{7.29}\%* & +\textbf{22.52}\%* & +\textbf{13.55}\%*
& +\textbf{9.02}\%* & +\textbf{9.99}\%* & +\textbf{9.94}\%* & +\textbf{10.65}\%* \\

\midrule
Pair-wise gen. + ToTER
& +2.04\% & +0.22\% & +3.80\% & +1.75\%
& +5.35\% & +6.63\% & +5.58\% & +6.53\% \\
\bottomrule
\end{tabular}}
\label{tab:compatibility}
\end{table*}

\subsubsection{\textbf{Compatibility with retrieval enhancement methods}}
Many recent methods enhance retrieval by incorporating additional signals such as LLM-generated semantic cues~\cite{mackie2023generative} and topic-level relevance modeling~\cite{ToTER}.
To examine the compatibility of \proposed with these approaches, we integrate it with two representative enhancement techniques:
(1) \textbf{GRF}~\cite{mackie2023generative}, which augments queries using LLM-generated contexts;
(2) \textbf{ToTER}~\cite{ToTER}, which measures relevance using taxonomy-grounded topic distributions.
In addition, the concept-aware relevance modeling in Eq.~\ref{eq:dual_score}, originally used in \proposed for query filtering, can also be applied at inference time to provide complementary concept-matching signals. We denote this extended variant as \proposed++.

Across all comparisons, we use the same retriever architecture (\ctr) fine-tuned with \proposed (prompting scheme: pair-wise generation).
Table~\ref{tab:compatibility} reports the performance gains.
The results show that our concept-based framework integrates well with existing enhancement strategies and consistently yields additional improvements.
Also, \proposed++ achieves the strongest overall performance across datasets, showing that concept-level signals offer complementary information beyond what conventional enhancement methods capture.
These findings indicate that the academic concept index can function as a flexible and broadly applicable component within diverse retrieval pipelines. 
Exploring integration with other techniques is a promising direction for future work.

\subsubsection{\textbf{Effectiveness of concept-based filtering}}
\begin{figure}[t]
\centering
\includegraphics[height=2.3cm]{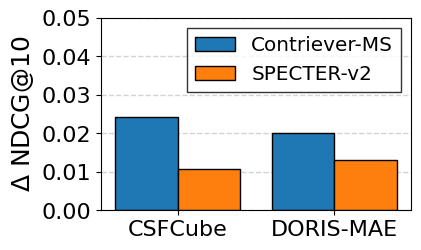}
\includegraphics[height=2.3cm]{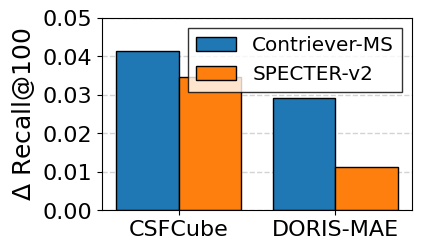}\hspace{-0.1cm}
\caption{Improvements by concept coverage-based filtering.}
\label{fig:filtering}
\end{figure}

Figure~\ref{fig:filtering} presents the improvements achieved through the filtering step, which aims to remove low-quality queries that the document does not answer.
The proposed filtering technique can enhance retrieval accuracy by incorporating concept information.
This enhanced accuracy helps to accurately measure round-trip consistency, effectively improving the effects of fine-tuning.

\subsubsection{\textbf{Results with \rev{smaller} LLM}}
\label{result:Llama}
In Table~\ref{tab:Llama}, we explore the effectiveness of the proposed approach using \rev{a smaller} LLM, namely Llama-3-8B, with \ctr as the backbone model. 
Consistent with the previous results, the proposed techniques (\proposed and \proposed++) consistently improve the existing method.
We expect \proposed to be effective with existing LLMs that possess a certain degree of capability. 
Since comparing different LLMs is not the focus of this work, we leave further investigation on more various LLMs and their comparison for future study.

\begin{table}[t]
\caption{Retrieval performance with Llama-3-8B. 
We report improvements over no Fine-Tune. $^*$ denotes $p < 0.05$ from paired t-test with pair-wise generation.}
\centering
\renewcommand{\arraystretch}{1}
\resizebox{0.6\linewidth}{!}{
\begin{tabular}{l l lll} \toprule
\textbf{Dataset} & \textbf{Method} & \textbf{N@10} & \textbf{N@20} & \textbf{R@100} \\ \midrule
\multirow{3}{*}{CSFCube} & Pair-wise generation & +5.25\% & +0.94\% & \textcolor{red}{- 0.21\%} \\
 & w/ CCQGen & +6.55\% & +7.82\%$^*$ & +5.48\%$^*$ \\
 & w/ CCQGen++ & +27.92\%$^*$ & +20.09\%$^*$ & +9.01\%$^*$ \\ \midrule
\multirow{3}{*}{DORIS-MAE} & Pair-wise generation & +0.00\% & +2.92\% & +5.43\% \\
 & w/ CCQGen & +5.19\%$^*$ & +9.57\%$^*$ & +6.69\% \\
 & w/ CCQGen++ & +16.75\%$^*$ & +20.87\%$^*$ & +14.65\%$^*$\\ \bottomrule
\end{tabular}}
\label{tab:Llama}
\end{table}

\input{sections/99992query_analysis}

\subsubsection{\textbf{Analysis of generated queries}}
\label{result:query_analysis}
We analyze whether \proposed indeed reduces redundancy among the queries and includes a variety of related terms.
We introduce two criteria: (1) \textbf{redundancy}, measured as the average cosine similarity of term frequency vectors of queries.\footnote{We use CountVectorizer from the SciKit-Learn library.}
A high redundancy indicates that queries tend to cover similar aspects of the document.
(2) \textbf{lexical overlap}, measured as the average BM25 score between the queries and the document.
A higher lexical overlap indicates that queries tend to reuse~terms~from~the~document.

In Table~\ref{tab:query_analysis}, the generated queries show higher lexical overlap with the document compared to the \rev{user queries}.
This shows that the generated queries tend to use a limited range of terms already present in the document, whereas user queries include a broader variety of terms.
With the ‘diverse condition’ (i.e., Promptgator\_diverse), the generated queries exhibit reduced lexical overlap and redundancy. 
However, this does not consistently lead to performance improvements.
The improved term usage often appears in common expressions, not necessarily enhancing concept coverage.
Conversely, \proposed directly guides each new query to complement the previous ones.
Also, \proposed incorporate concept-related terms not explicitly mentioned in the document via enrichment step (Section \ref{method:enrich}).
This provides more systematic controls over the generation, leading to consistent improvements.
\rev{Examples of generated queries comparing \proposed with baselines, along with a failure case, are provided in Appendix~\ref{appendix:examples}.}

\subsubsection{\rev{\textbf{Robustness to Taxonomy Quality}}}
{\rev{
We further analyze the robustness of CCQGen to taxonomy quality. 
We examine two aspects: (1) \textbf{Topical coverage}, which reflects how comprehensively the taxonomy covers diverse academic topics.
To simulate limited coverage, we randomly prune a fraction of taxonomy nodes and re-run \proposed on the remaining nodes.
(2) \textbf{Taxonomy signal precision}, which reflects the accuracy of topical signals derived from the taxonomy.
We control this by varying the proportion of documents to which LLM-based topic selection is applied.
We use \ctr\ as the backbone retriever on \csfcube dataset.}

\begin{figure*}[t]
    \centering
    \begin{subfigure}[b]{0.7\linewidth}
        \centering
        \includegraphics[width=\linewidth]{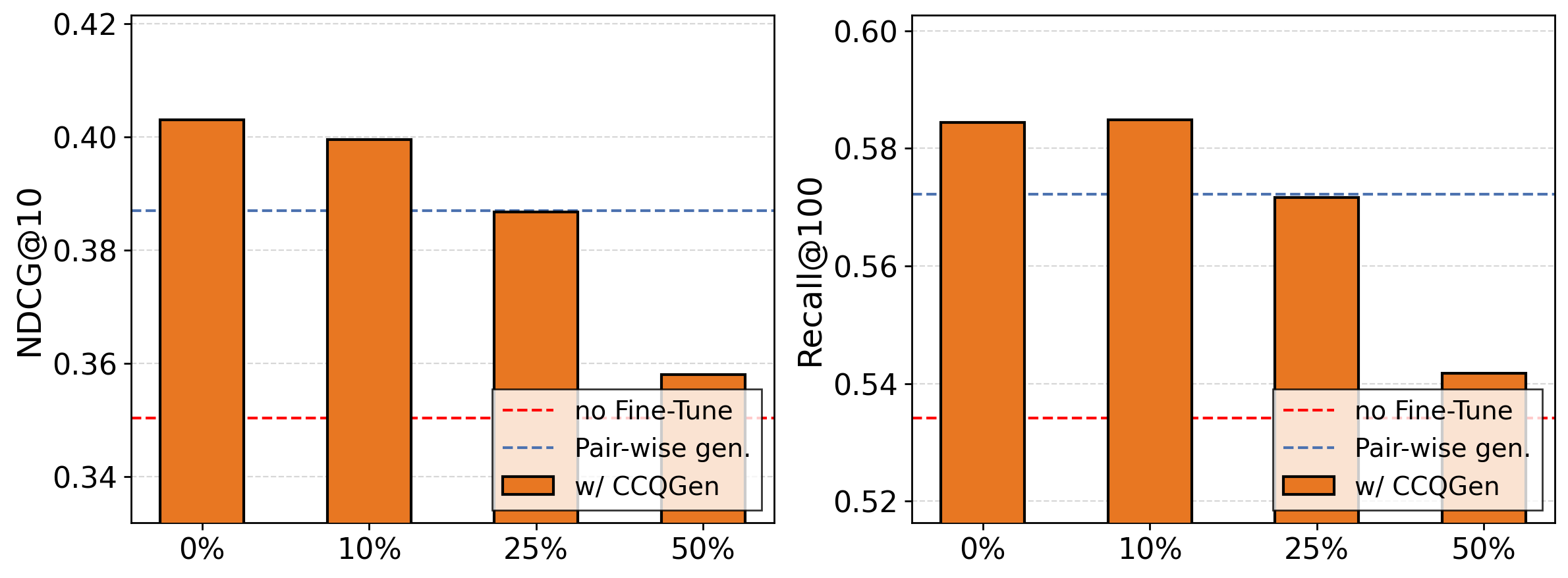}
        \caption{\rev{Effect of topical coverage}}
        \label{fig:taxonomy_ablation_coverage}
    \end{subfigure}
    \\[0.5em]
    \begin{subfigure}[b]{0.7\linewidth}
        \centering
        \includegraphics[width=\linewidth]{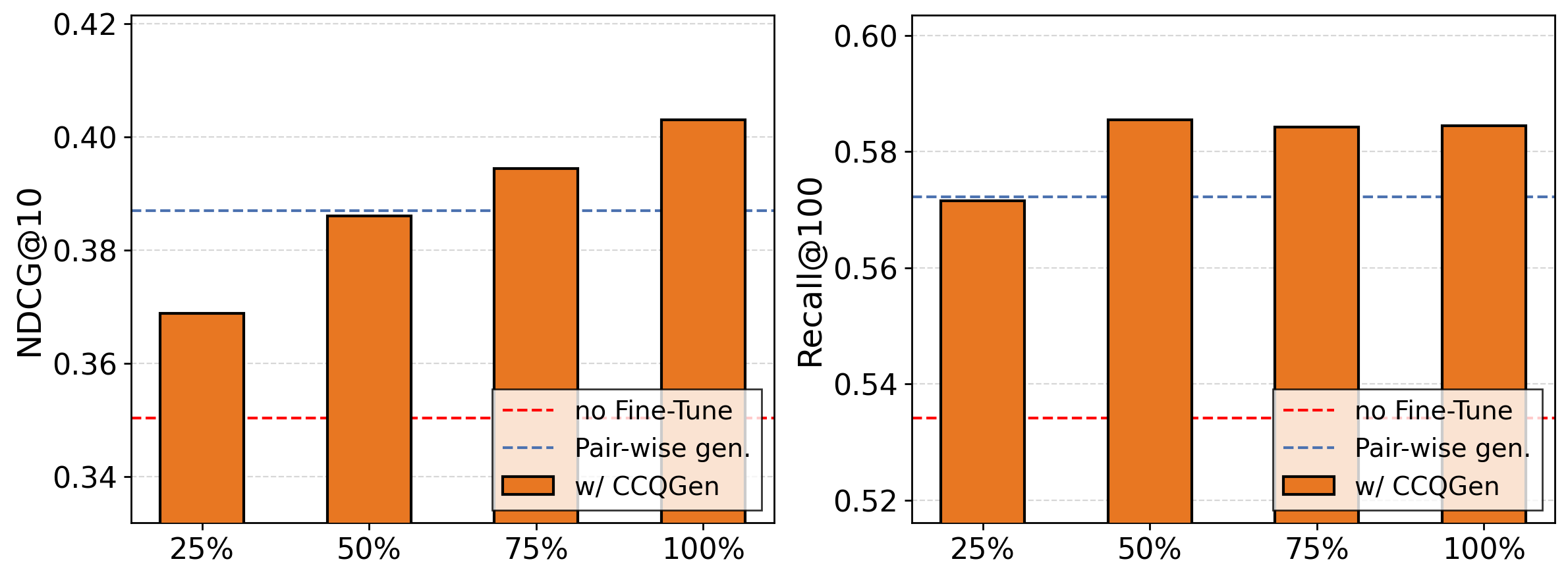}
        \caption{\rev{Effect of taxonomy signal precision}}
        \label{fig:taxonomy_ablation_filtering}
    \end{subfigure}
    \caption{
    \rev{Analysis on taxonomy quality.
    (a) shows the effect of pruning taxonomy nodes, and (b) shows the effect of varying the LLM-based filtering ratio.
Dashed lines indicate the no fine-tune (red) and Pair-wise generation (blue) baselines.}
    }
    \label{fig:taxonomy_ablation}
\end{figure*}

\smallsection{\rev{Topical coverage}}
\rev{
As shown in Figure~\ref{fig:taxonomy_ablation_coverage}, performance degrades as more taxonomy nodes are pruned, as broader coverage enables more accurate concept identification.
However, \proposed consistently outperforms both the no fine-tune and Pair-wise generation baselines across all pruning ratios up to 25\%, and remains competitive even under 50\% pruning.
This indicates that \proposed provides reliable improvements even when the taxonomy is incomplete.
Overall, our method does not rely on perfect taxonomy coverage and remains effective under moderate incompleteness, making it applicable in domains where comprehensive taxonomies are unavailable.
}}

\smallsection{\rev{Taxonomy signal precision}}
\rev{
To examine sensitivity to taxonomy signal precision, we vary the proportion of documents that undergo LLM-based topic selection.
A lower ratio reduces the accuracy of the resulting topical signals, while a ratio of 100\% corresponds to the full setting used in \proposed.
As shown in Figure~\ref{fig:taxonomy_ablation_filtering}, performance improves as the ratio increases and stabilizes near the full setting.
Up to a 50\% filtering ratio, \proposed achieves comparable or higher performance than the Pair-wise generation baseline.
Even at a low ratio of 25\%, it still provides consistent improvements over the backbone (no fine-tune) model.}

\rev{
Together, these results confirm that \proposed is resilient to realistic imperfections in taxonomy quality.
Furthermore, it is worth noting that automatic taxonomy construction and completion are well-established research areas with many readily available tools~\cite{lee2022taxocom, shi2024taxonomy}, which can be applied to obtain high-quality taxonomies.
}

%% file: sections/99991main_table.tex
\begin{table*}[t]
\caption{Performance comparison after fine-tuning with the generated queries. \textcolor{red}{Red} color denotes results that fail to show improvements over no Fine-Tune. $^{\dagger}$ and * indicate a statistically significant difference ($p<0.05$) from no Fine-Tune (one-sample t-test) and the applied query generation method (paired t-test), respectively.}
\centering
\renewcommand{\arraystretch}{1}
\resizebox{\linewidth}{!}{
\begin{tabular}{cl llllll  llllll}\toprule
& \multicolumn{1}{c}{\multirow{2}{*}{\textbf{Query generation}}} & \multicolumn{6}{c}{\textbf{CSFCube}} & \multicolumn{6}{c}{\textbf{DORIS-MAE}} \\ \cmidrule(lr){3-8} \cmidrule(lr){9-14}
 &  & \textbf{N@10} & \textbf{N@20} & \textbf{M@10} & \textbf{M@20} & \textbf{R@50} & \textbf{R@100} & \textbf{N@10} & \textbf{N@20} & \textbf{M@10} & \textbf{M@20} & \textbf{R@50} & \textbf{R@100} \\ \midrule
\parbox[t]{2mm}{\multirow{7}{*}{\rotatebox[origin=c]{90}{\ctr}}} & no Fine-Tune & 0.3313 & 0.3604 & 0.1525 & 0.1937 & 0.5783 & 0.7136 & 0.2603 & 0.2707 & 0.1177 & 0.1422 & 0.4509 & 0.5877 \\
 & GenQ & 0.3401 & \textcolor{red}{0.3495} & \textcolor{red}{0.1476} & \textcolor{red}{0.1841} & \textcolor{red}{0.5571} & \textcolor{red}{0.6843} & \textcolor{red}{0.2496} & \textcolor{red}{0.2647} & \textcolor{red}{0.1152} & \textcolor{red}{0.1396} & 0.4598 & \textcolor{red}{0.5805} \\
 & Promptgator\_{diverse} & 0.3539 & 0.3771 & 0.1606 & 0.2029 & 0.5950 & \textcolor{red}{0.7132} & \textcolor{red}{0.2461} & \textcolor{red}{0.2690} & \textcolor{red}{0.1143} & \textcolor{red}{0.1406} & 0.4645 & 0.5951 \\ \cmidrule(lr){2-14}
 & Promptgator & 0.3441 & 0.3670 & 0.1538 & 0.1974 & 0.5928 & 0.7298 & \textcolor{red}{0.2526} & 0.2724 & \textcolor{red}{0.1161} & \textcolor{red}{0.1418} & 0.4718 & 0.5961 \\
 & w/ CCQGen (ours) & \textbf{0.3605}$^{\dagger}$ & \textbf{0.3991}$^{\dagger}$$^*$ & \textbf{0.1614}$^{\dagger}$ & \textbf{0.2194}$^{\dagger}$$^*$ & \textbf{0.6333}$^{\dagger}$$^*$ & \textbf{0.7467}$^{\dagger}$ & \textbf{0.2697}$^*$ & \textbf{0.2883}$^{\dagger}$$^*$ & \textbf{0.1267}$^{\dagger}$$^*$ & \textbf{0.1536}$^{\dagger}$$^*$ & \textbf{0.4983}$^{\dagger}$$^*$ & \textbf{0.6327}$^{\dagger}$$^*$ \\ \cmidrule(lr){2-14}
 & Pair-wise generation & 0.3418 & 0.3686 & \textcolor{red}{0.1522} & 0.1971 & 0.5961 & 0.7225 & \textcolor{red}{0.2541} & 0.2753 & \textcolor{red}{0.1177} & 0.1445 & 0.4809 & 0.5947 \\
 & w/ CCQGen (ours) & \textbf{0.3670}$^{\dagger}$$^*$ & \textbf{0.4063}$^{\dagger}$$^*$ & \textbf{0.1656}$^{\dagger}$$^*$ & \textbf{0.2228}$^{\dagger}$$^*$ & \textbf{0.6362}$^{\dagger}$$^*$ & \textbf{0.7526}$^{\dagger}$$^*$ & \textbf{0.2783}$^{\dagger}$$^*$ & \textbf{0.2943}$^{\dagger}$$^*$ & \textbf{0.1308}$^{\dagger}$$^*$ & \textbf{0.1577}$^{\dagger}$$^*$ & \textbf{0.5089}$^{\dagger}$$^*$ & \textbf{0.6331}$^{\dagger}$$^*$ \\ \midrule
\parbox[t]{2mm}{\multirow{7}{*}{\rotatebox[origin=c]{90}{SPECTER-v2}}} & no Fine-Tune & 0.3503 & 0.3579 & 0.1615 & 0.2043 & 0.5341 & 0.6859 & 0.2121 & 0.2283 & 0.0942 & 0.1147 & 0.4182 & 0.5441 \\
 & GenQ & 0.3658 & 0.3659 & 0.1699 & 0.2083 & 0.5541 & \textcolor{red}{0.6836} & 0.2338 & 0.2525 & 0.1045 & 0.1287 & 0.4412 & 0.5613 \\
 & Promptgator\_diverse & 0.3672 & 0.3801 & 0.1721 & 0.2157 & 0.5687 & 0.6972 & 0.2469 & 0.2733 & 0.1121 & 0.1401 & 0.4843 & 0.6102 \\ \cmidrule(lr){2-14}
 & Promptgator & 0.3766 & 0.3886 & 0.1790 & 0.2245 & 0.5715 & 0.6962 & 0.2479 & 0.2713 & 0.1131 & 0.1398 & 0.4851 & 0.6064 \\
 & w/ CCQGen (ours) & \textbf{0.4105}$^{\dagger}$$^*$ & \textbf{0.4176}$^{\dagger}$$^*$ & \textbf{0.2085}$^{\dagger}$$^*$ & \textbf{0.2549}$^{\dagger}$$^*$ & \textbf{0.5886}$^{\dagger}$ & \textbf{0.7355}$^{\dagger}$$^*$ & \textbf{0.2634}$^{\dagger}$$^*$ & \textbf{0.2891}$^{\dagger}$$^*$ & \textbf{0.1226}$^{\dagger}$ & \textbf{0.1520}$^{\dagger}$$^*$ & \textbf{0.4988}$^{\dagger}$ & \textbf{0.6265}$^{\dagger}$ \\ \cmidrule(lr){2-14}
 & Pair-wise generation & 0.3870 & 0.3999 & 0.1966 & 0.2423 & 0.5722 & 0.6972 & 0.2523 & 0.2782 & 0.1163 & 0.1442 & 0.4885 & 0.6148 \\
 & w/ CCQGen (ours) & \textbf{0.4031}$^{\dagger}$$^*$ & \textbf{0.4150}$^{\dagger}$ & \textbf{0.2040}$^{\dagger}$ & \textbf{0.2534}$^{\dagger}$$^*$ & \textbf{0.5844}$^{\dagger}$ & \textbf{0.7333}$^{\dagger}$$^*$ & \textbf{0.2681}$^{\dagger}$$^*$ & \textbf{0.2932}$^{\dagger}$$^*$ & \textbf{0.1247}$^{\dagger}$ & \textbf{0.1546}$^{\dagger}$$^*$ & \textbf{0.5064}$^{\dagger}$ & \textbf{0.6304}$^{\dagger}$\\ \bottomrule
\end{tabular}}
\label{tab:main}
\end{table*}

%% file: sections/99992query_analysis.tex
\begin{table}[t]
\caption{Analysis of generated queries.
(a) Statistics of queries generated by different methods.
(b) Retriever performance (\specter on NDCG@10) after fine-tuning using the queries.
The average lexical overlap of \rev{user queries (i.e., the test queries provided with each dataset)} is $13.32$ for \csfcube and $20.42$ for \dorismae.}
\centering
\resizebox{0.7\linewidth}{!}{
\begin{tabular}{l l l l l}
\toprule
\multirow{2}{*}{\textbf{Dataset}}  &
\multirow{2}{*}{\textbf{Query generation}} 
& \multicolumn{2}{c}{\textbf{(a) Query statistics}} 
& \multirow{2}{*}{\parbox{2.2cm}{\centering \textbf{(b) Retriever} \\ \textbf{performance}}} \\
\cmidrule(lr){3-4}
&  & \textbf{redundancy ($\downarrow$)} & \textbf{lexical overlap ($\downarrow$)} &  \\ 
\midrule
& Promptgator 
& 0.5072 
& 31.51 
& 0.3766 \\
\textbf{CSFCube} 
& w/ diverse condition 
& 0.4512 (-11.0\%) 
& \textbf{24.05 (-23.7\%)} 
& 0.3672 (-2.6\%) \\
& w/ \proposed 
& \textbf{0.3997 (-21.2\%)} 
& 24.41 (-22.5\%) 
& \textbf{0.4105 (+9.0\%)} \\
\midrule
& Promptgator 
& 0.4861 
& 53.58 
& 0.2479 \\
\textbf{DORIS-MAE} 
& w/ diverse condition 
& \textbf{0.3958 (-18.6\%)} 
& 41.56 (-22.4\%) 
& 0.2469 (-0.4\%) \\
& w/ \proposed 
& 0.3993 (-17.9\%) 
& \textbf{40.54 (-24.3\%)} 
& \textbf{0.2634 (+6.2\%)} \\
\bottomrule
\end{tabular}}
\label{tab:query_analysis}
\end{table}

%% file: sections/052Experiment_result2.tex
\subsection{Experiment results for \proposedtwo}
\label{tab:CCExpand_results}
In this section, we present comprehensive experimental evaluations of \proposedtwo.

\subsubsection{\textbf{Compared methods}} 
We compare \proposedtwo with a range of LLM-based auxiliary context generation approaches.
First, we consider recent methods that expand either the document or the query using LLM-generated contexts:
\begin{itemize}[leftmargin=*]
\item \textbf{Doc expansion}: A widely used strategy~\cite{doc2query, doc2querymm, nogueira2019doc2query} is to augment each document with synthetic queries to mitigate term mismatch with real user queries.
Following prior work, we evaluate two variants of document expansion:  
(i) the strongest competing query generation method (i.e., pair-wise generation), denoted \textbf{BestQGen}, and  
(ii) document expansion using our concept-aware queries generated by \proposed, denoted \textbf{Ours}.

\item \textbf{GRF}~\cite{mackie2023generative}: It performs query expansion by prompting LLMs to generate diverse auxiliary contexts such as keywords, entities, and summaries.  
The expanded queries are directly used during relevance matching.

\item \textbf{\proposedtwo}$--$: A variant of \proposedtwo that averages all LLM-generated snippets into a single vector representation.  
This baseline isolates the benefit of selecting the most concept-aligned snippet, allowing us to measure the importance of focused concept matching.
\end{itemize}
Detailed comparative analysis with these methods is provided in Section~\ref{subsubsec:CCExpand_result1}.
Second, we focus on comparing with \textbf{HyDE}~\cite{hyde}, a recent context generation method closely related to \proposedtwo.
\begin{itemize}[leftmargin=*]
\item \textbf{HyDE}: It generates \textit{multiple hypothetical documents} conditioned on the query. 
These documents introduce terms and expression styles not explicitly present in the original query, enriching the semantics for retrieval. 
The final relevance is computed based on the \textit{weighted sum} of similarities with both the original query and the generated hypothetical documents.
\end{itemize}
Though effective, it does not consider document-level concept structure and performs online expansion, which increases latency at inference time.
Furthermore, because HyDE improves query-side context, it can be combined with our document-side augmentation (\proposedtwo) for additional benefits.
The detailed results are presented in Section~\ref{subsubsec:CCExpand_result2}.


\input{./sections/99993_CCExpand}

\subsubsection{\textbf{Comparison with document and query expansion methods}}
\label{subsubsec:CCExpand_result1}
Table~\ref{tab:cc_expand_results} presents retrieval performances with various context generation methods.
\proposedtwo outperforms all compared methods across most metrics. 
Notably, simply expanding documents~(i.e., Doc exp. (BestQGen)) or queries (i.e., GRF) often degrades the retrieval performance, indicating that naively leveraging LLM-generated contexts does not necessarily guarantee improvements.
Interestingly, `Doc exp. (Ours)', a concatenation of our concept-aware queries to documents, yields consistent gains compared to `Doc exp. (BestQGen)'.
This underscores the importance of concept-grounded signals in context augmentation. 
However, this approach remains limited: once the expanded document is compressed into a single embedding, much of its \rev{concept-aware} information becomes diluted.

\rev{
Interestingly, on CSFCube with \ctr, Doc exp.\ (Ours) achieves comparable or even higher NDCG and MAP than \proposedtwo. This highlights the quality of CCQGen queries, as even simple augmentation through direct concatenation yields meaningful improvements.
However, overall, \proposedtwo achieves higher performance across diverse backbones and datasets.}
\proposedtwo achieves such improvements by explicitly identifying and utilizing the snippet most aligned with the query.
Notably, averaging similarity scores over all snippets (i.e., \proposedtwo$--$) yields limited effectiveness, as the averaging dilutes the most discriminative concept signals.
This result supports our design choice of selecting the most relevant snippet for relevance~matching.



\subsubsection{\textbf{Comparison with HyDE}}
\label{subsubsec:CCExpand_result2}
We provide a comparative analysis with HyDE in terms of both retrieval effectiveness and efficiency, highlighting the advantages of our concept-focused expansion.


\input{./sections/99994_CCExpand_HyDE}
\smallsection{Retrieval effectiveness} 
Table~\ref{tab:hyde_cc_expand_results} summarizes the retrieval performance. 
For a fair comparison, both HyDE and \proposedtwo generate the same number of hypothetical documents and snippets (five in our experiments).
The results reveal two key findings regarding stability and synergy.
First, HyDE shows notable performance instability across different backbones; with \ctr, it often underperforms the backbone itself (marked in red), likely due to hallucinations or domain mismatches introducing noise. 
In contrast, \proposedtwo consistently outperforms the backbone in all settings, showing strong robustness. 
\rev{On the other hand, the hybrid approach, HyDE + \proposedtwo, achieves competitive overall performance, particularly with \specter}.
This suggests a complementary relationship:
while HyDE enriches the query context, \proposedtwo anchors the relevance matching with precise, concept-grounded snippets from the document, which can mitigate noise introduced by HyDE.

\rev{However, we note that on \ctr, \proposedtwo alone outperforms HyDE + \proposedtwo on most metrics.
We conjecture that HyDE may degrade performance with \ctr, potentially due to noise from hypothetical documents that are not well aligned with the target corpus, which interferes with the concept-grounded matching of \proposedtwo.
In contrast, with \specter, the combination yields consistent gains.
We conjecture that this is because \specter, as a model specialized for scientific domains, may encode HyDE-generated documents more effectively, leading to better alignment between the generated contexts and document representations.}

\input{./sections/99995_latency}
\smallsection{Effectiveness-efficiency trade-off} 
In addition to retrieval performance, inference latency is a critical factor for real-world deployment. 
Here, we investigate the trade-off between effectiveness and inference latency. 
Latency is measured on an Intel Xeon Gold 6530 CPU with a single NVIDIA RTX PRO 6000 Blackwell Max-Q GPU.
While HyDE yields improvements to some extent, these gains come at the costs of substantially higher latency.
This reflects its core limitation: the reliance on LLM inference at serving time, making it difficult to deploy in latency-sensitive retrieval systems.

\rev{Figure~\ref{fig:acc_latency} highlights this contrast: HyDE generates hypothetical documents online at query time, incurring LLM inference latency for each query, whereas \proposedtwo performs all snippet generation and indexing offline, requiring only efficient vector similarity computation during retrieval.
As a result, \proposedtwo achieves approximately 40--100$\times$ lower inference latency than HyDE, while consistently delivering stronger retrieval performance.}



\input{./sections/99996_m_diversity}
\subsubsection{\textbf{Hyperparameter study}} 
\label{subsubsec:CCExpand_study}
To guide the hyperparameter selection of \proposedtwo, we conduct a detailed analysis on DORIS-MAE using \specter as the backbone model. 
In Figure~\ref{fig:parameter_anaylsis}, we examine two hyperparameters: the balancing coefficient $\alpha$, which controls the contribution of concept-focused similarity, and the number of snippets $M$.


\smallsection{Impact of the balancing coefficient} 
Most metrics peak at values between 0 and 1, and the best results are obtained around $\alpha \in \{0.6, 0.8, 1.0\}$, indicating that the original document and the concept-focused snippets provide complementary semantic signals. 
This observation further supports our methodology, showing that representing both texts as separate vectors enables more effective relevance matching between user queries and scientific documents.


\smallsection{Impact of the number of snippets} 
We observe a substantial performance leap when increasing $M$ from 0 to 5, where $M=0$ corresponds to using only the original document.
The performance gains become less pronounced after $M=3$, and the best results are obtained at $M \in \{4,5\}$, confirming the benefit of incorporating multiple \rev{concept-aware} views of the document.
Thus, while larger $M$ yields the highest effectiveness, using $M=3$ offers a practical trade-off by reducing generation cost while retaining most of the performance gains.




\input{sections/99998_case}

\subsubsection{\textbf{\rev{Examples} of \proposedtwo}}
\rev
{Table~\ref{tab:rank_improvement_case} presents concrete examples of how \proposedtwo improves document ranking on DORIS-MAE.
By tailoring each snippet as a direct response to the concept-aware query, the generated text elaborates the underlying concepts using broader and related terminology, yielding higher semantic overlap with the user query. 
In both cases, the relevant document is initially ranked below position 50 and thus missed by the backbone retriever.
With \proposedtwo, the relevant document is promoted to a substantially higher rank (rank 2 and rank 5, respectively) by incorporating the most concept-aligned snippet during relevance matching. 
This illustrates how snippet-level concept matching recovers documents that are diluted in the single-vector representation.
Additional examples are provided in Appendix~\ref{appendix:examples}.
}



%% file: sections/99993_CCExpand.tex
\begin{table*}[t]
\caption{Retrieval performance comparison with various context-augmentation methods. All methods are training-free and applied to the same retrieval model (denoted Retriever). \textcolor{red}{Red} color denotes results that fail to show improvements over Retriever. \rev{$^\dagger$ indicates a statistically significant difference ($p < 0.05$) from Retriever (one-sample t-test).}}
\centering
\resizebox{\linewidth}{!}{
\begin{tabular}{cl llllll llllll}\toprule
& \multicolumn{1}{c}{\multirow{2}{*}{\textbf{Method}}} & \multicolumn{6}{c}{\textbf{CSFCube}} & \multicolumn{6}{c}{\textbf{DORIS-MAE}} \\ \cmidrule(lr){3-8} \cmidrule(lr){9-14}
 &  & \textbf{N@10} & \textbf{N@20} & \textbf{M@10} & \textbf{M@20} & \textbf{R@50} & \textbf{R@100} & \textbf{N@10} & \textbf{N@20} & \textbf{M@10} & \textbf{M@20} & \textbf{R@50} & \textbf{R@100} \\ \midrule

\parbox[t]{2mm}{\multirow{6}{*}{\rotatebox[origin=c]{90}{\ctr}}}
 & Retriever
   & 0.3313 & 0.3604 & 0.1525 & 0.1937 & 0.5783 & 0.7136
   & 0.2603 & 0.2707 & 0.1177 & 0.1422 & 0.4509 & 0.5877 \\
 & GRF
   & \textcolor{red}{0.3047} & \textcolor{red}{0.3321} & \textcolor{red}{0.1388} & \textcolor{red}{0.1780} & \textcolor{red}{0.5576} & \textcolor{red}{0.6697}
   & \textcolor{red}{0.2538} & 0.2712 & \textcolor{red}{0.1134} & \textcolor{red}{0.1393} & 0.4602\rev{$^\dagger$} & 0.5950 \\
 & Doc exp. (BestQGen)
   & 0.3341 & 0.3631 & \textcolor{red}{0.1511} & 0.1948 & \textcolor{red}{0.5689} & \textcolor{red}{0.6868}
   & \textcolor{red}{0.2403} & \textcolor{red}{0.2505} & \textcolor{red}{0.1059} & \textcolor{red}{0.1277} & \textcolor{red}{0.4280} & \textcolor{red}{0.5554} \\
 & Doc exp. (Ours)
   & \textbf{0.3498}\rev{$^\dagger$} & \textbf{0.3759}\rev{$^\dagger$} & \textbf{0.1615}\rev{$^\dagger$} & \textbf{0.2074}\rev{$^\dagger$} & 0.5809 & \textcolor{red}{0.7046}
   & \textcolor{red}{0.2483} & \textcolor{red}{0.2662} & \textcolor{red}{0.1120} & \textcolor{red}{0.1378} & 0.4569 & 0.5963 \\
 & \rev{CCExpand$^{--}$} (Ours)
   & \textcolor{red}{0.3268} & \textcolor{red}{0.3513} & 0.1532 & \textcolor{red}{0.1904} & 0.5829 & \textcolor{red}{0.7105}
   & 0.2712\rev{$^\dagger$} & 0.2810\rev{$^\dagger$} & 0.1253\rev{$^\dagger$} & 0.1503\rev{$^\dagger$} & 0.4755\rev{$^\dagger$} & 0.6114\rev{$^\dagger$} \\
 & \textbf{CCExpand (Ours)}
   & 0.3412\rev{$^\dagger$} & 0.3708\rev{$^\dagger$} & 0.1579\rev{$^\dagger$} & 0.2001\rev{$^\dagger$} & \textbf{0.6008}\rev{$^\dagger$} & \textbf{0.7141}
   & \textbf{0.2730}\rev{$^\dagger$} & \textbf{0.2871}\rev{$^\dagger$} & \textbf{0.1266}\rev{$^\dagger$} & \textbf{0.1530}\rev{$^\dagger$} & \textbf{0.4802}\rev{$^\dagger$} & \textbf{0.6155}\rev{$^\dagger$} \\ \midrule

\parbox[t]{2mm}{\multirow{6}{*}{\rotatebox[origin=c]{90}{\specter}}}
 & Retriever
   & 0.3503 & 0.3579 & 0.1615 & 0.2043 & 0.5341 & 0.6859
   & 0.2121 & 0.2283 & 0.0942 & 0.1147 & 0.4182 & 0.5441 \\
 & GRF
   & \textcolor{red}{0.3075} & \textcolor{red}{0.3206} & \textcolor{red}{0.1389} & \textcolor{red}{0.1787} & \textcolor{red}{0.5325} & \textcolor{red}{0.6495}
   & \textcolor{red}{0.2029} & \textcolor{red}{0.2265} & \textcolor{red}{0.0903} & \textcolor{red}{0.1122} & 0.4394\rev{$^\dagger$} & 0.5689\rev{$^\dagger$} \\
 & Doc exp. (BestQGen)
   & \textcolor{red}{0.2958} & \textcolor{red}{0.3039} & \textcolor{red}{0.1176} & \textcolor{red}{0.1507} & 0.5342 & \textcolor{red}{0.6509}
   & \textcolor{red}{0.2016} & \textcolor{red}{0.2146} & \textcolor{red}{0.0831} & \textcolor{red}{0.1018} & \textcolor{red}{0.4049} & \textcolor{red}{0.5402} \\
 & Doc exp. (Ours)
   & \textcolor{red}{0.3427} & \textcolor{red}{0.3567} & \textcolor{red}{0.1530} & \textcolor{red}{0.1997} & 0.5346 & \textcolor{red}{0.6854}
   & \textcolor{red}{0.2052} & \textcolor{red}{0.2227} & \textcolor{red}{0.0877} & \textcolor{red}{0.1090} & \textcolor{red}{0.4063} & \textcolor{red}{0.5330} \\
 & \rev{CCExpand$^{--}$} (Ours)
   & \textcolor{red}{0.3422} & 0.3639\rev{$^\dagger$} & \textcolor{red}{0.1553} & 0.2086\rev{$^\dagger$} & 0.5381 & \textbf{0.6999}\rev{$^\dagger$}
   & 0.2250\rev{$^\dagger$} & 0.2435\rev{$^\dagger$} & 0.0993\rev{$^\dagger$} & 0.1221\rev{$^\dagger$} & 0.4402\rev{$^\dagger$} & 0.5674\rev{$^\dagger$} \\
 & \textbf{CCExpand (Ours)}
   & \textbf{0.3641}\rev{$^\dagger$} & \textbf{0.3740}\rev{$^\dagger$} & \textbf{0.1639}\rev{$^\dagger$} & \textbf{0.2142}\rev{$^\dagger$} & \textbf{0.5571}\rev{$^\dagger$} & 0.6948\rev{$^\dagger$}
   & \textbf{0.2362}\rev{$^\dagger$} & \textbf{0.2517}\rev{$^\dagger$} & \textbf{0.1065}\rev{$^\dagger$} & \textbf{0.1285}\rev{$^\dagger$} & \textbf{0.4440}\rev{$^\dagger$} & \textbf{0.5744}\rev{$^\dagger$} \\ \bottomrule
\end{tabular}}
\label{tab:cc_expand_results}
\end{table*}

%% file: sections/99994_CCExpand_HyDE.tex

\begin{table*}[t]
\caption{Retrieval performance comparison with HyDE and CCExpand integration. The best performance for each metric is marked in \textbf{bold}. Values lower than the backbone retriever are marked in \textcolor{red}{red}.}
\centering
\renewcommand{\arraystretch}{1.2}
\resizebox{\linewidth}{!}{
\begin{tabular}{cl llllll llllll}\toprule
& \multicolumn{1}{c}{\multirow{2}{*}{\textbf{Method}}} & \multicolumn{6}{c}{\textbf{CSFCube}} & \multicolumn{6}{c}{\textbf{DORIS-MAE}} \\ \cmidrule(lr){3-8} \cmidrule(lr){9-14}
 &  & \textbf{N@10} & \textbf{N@20} & \textbf{M@10} & \textbf{M@20} & \textbf{R@50} & \textbf{R@100} & \textbf{N@10} & \textbf{N@20} & \textbf{M@10} & \textbf{M@20} & \textbf{R@50} & \textbf{R@100} \\ \midrule

\multirow{4}{*}{\rotatebox[origin=c]{90}{\small Contriever-MS}} 
 & Retriever & 0.3313 & 0.3604 & 0.1525 & 0.1937 & 0.5783 & 0.7136 & 0.2603 & 0.2707 & 0.1177 & 0.1422 & 0.4509 & 0.5877 \\
 & CCExpand (Ours) & 0.3412 & \textbf{0.3708} & 0.1579 & \textbf{0.2001} & \textbf{0.6008} & \textbf{0.7141} & \textbf{0.2730} & 0.2871 & \textbf{0.1266} & \textbf{0.1530} & 0.4802 & \textbf{0.6155} \\
 & HyDE & 0.3479 & \textcolor{red}{0.3461} & 0.1560 & \textcolor{red}{0.1909} & \textcolor{red}{0.5553} & \textcolor{red}{0.6873} & \textcolor{red}{0.2471} & \textcolor{red}{0.2654} & \textcolor{red}{0.1073} & \textcolor{red}{0.1342} & 0.4662 & 0.5990 \\ 
 & HyDE + CCExpand (Ours) & \textbf{0.3537} & \textcolor{red}{0.3557} & \textbf{0.1600} & 0.1959 & \textcolor{red}{0.5667} & \textcolor{red}{0.6977} & 0.2641 & \textbf{0.2886} & 0.1210 & 0.1511 & \textbf{0.4898} & 0.6125 \\ \midrule

\multirow{4}{*}{\rotatebox[origin=c]{90}{\small SPECTER-v2}} 
 & Retriever & 0.3503 & 0.3579 & 0.1615 & 0.2043 & 0.5341 & 0.6859 & 0.2121 & 0.2283 & 0.0942 & 0.1147 & 0.4182 & 0.5441 \\
 & CCExpand (Ours) & 0.3641 & 0.3740 & 0.1639 & 0.2142 & 0.5571 & 0.6948 & 0.2362 & 0.2517 & 0.1065 & 0.1285 & 0.4440 & 0.5744 \\
 & HyDE & 0.3570 & 0.3703 & 0.1661 & 0.2146 & 0.5790 & 0.6924 & 0.2446 & 0.2678 & 0.1088 & 0.1365 & 0.4986 & 0.6083 \\ 
 & HyDE + CCExpand (Ours) & \textbf{0.3873} & \textbf{0.3913} & \textbf{0.1825} & \textbf{0.2313} & \textbf{0.5870} & \textbf{0.7212} & \textbf{0.2577} & \textbf{0.2807} & \textbf{0.1141} & \textbf{0.1437} & \textbf{0.5120} & \textbf{0.6288} \\ \bottomrule
\end{tabular}}
\label{tab:hyde_cc_expand_results}
\end{table*}

%% file: sections/99995_latency.tex
\begin{figure}[t]
    \centering
    \begin{subfigure}[b]{0.35\linewidth}
        \centering
        \includegraphics[height=3cm]{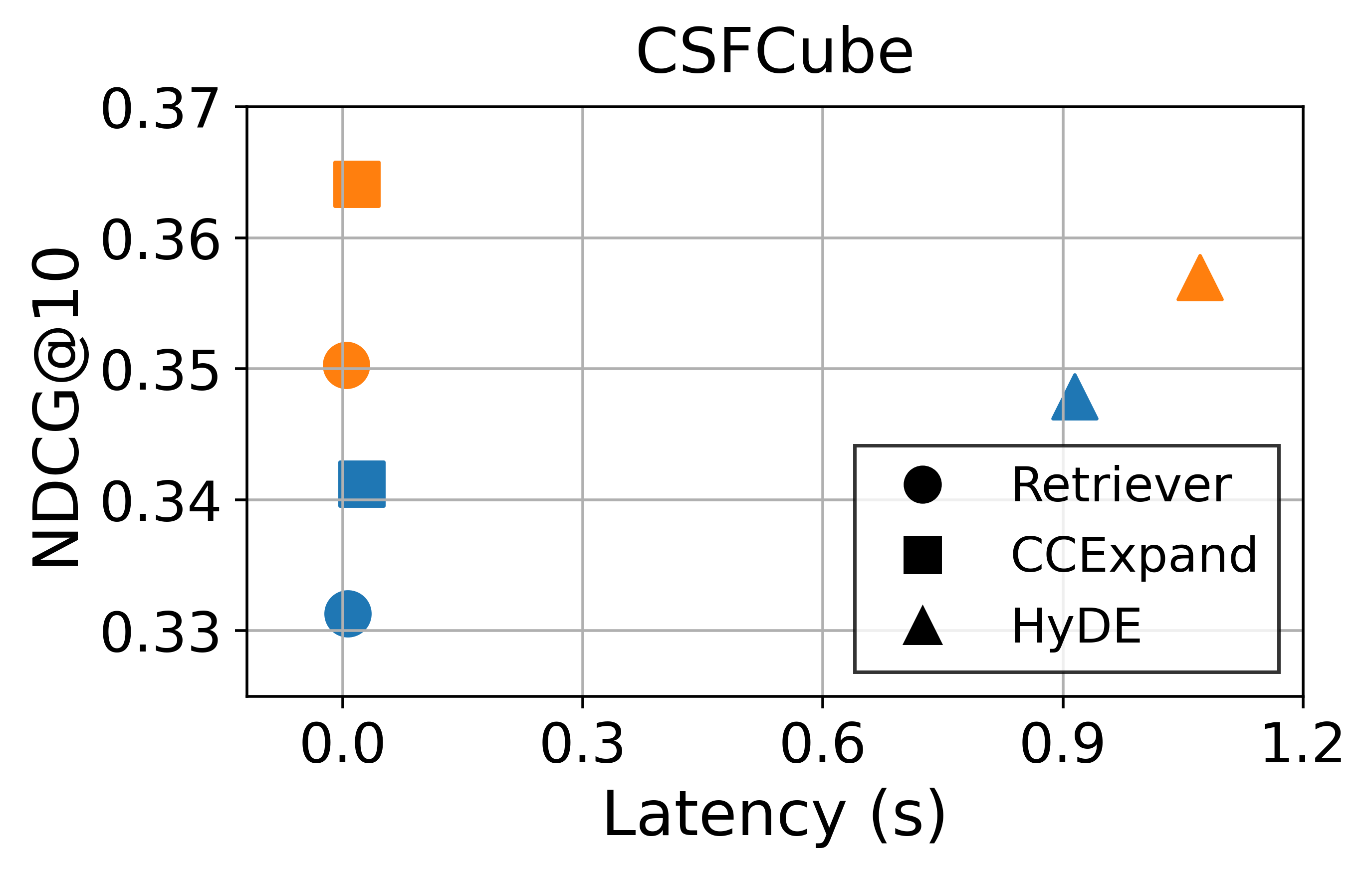}
        
        \label{fig:acc_latency_csfcube}
    \end{subfigure}
    \hspace{0.2cm}
    \begin{subfigure}[b]{0.35\linewidth}
        \centering
        \includegraphics[height=3cm]{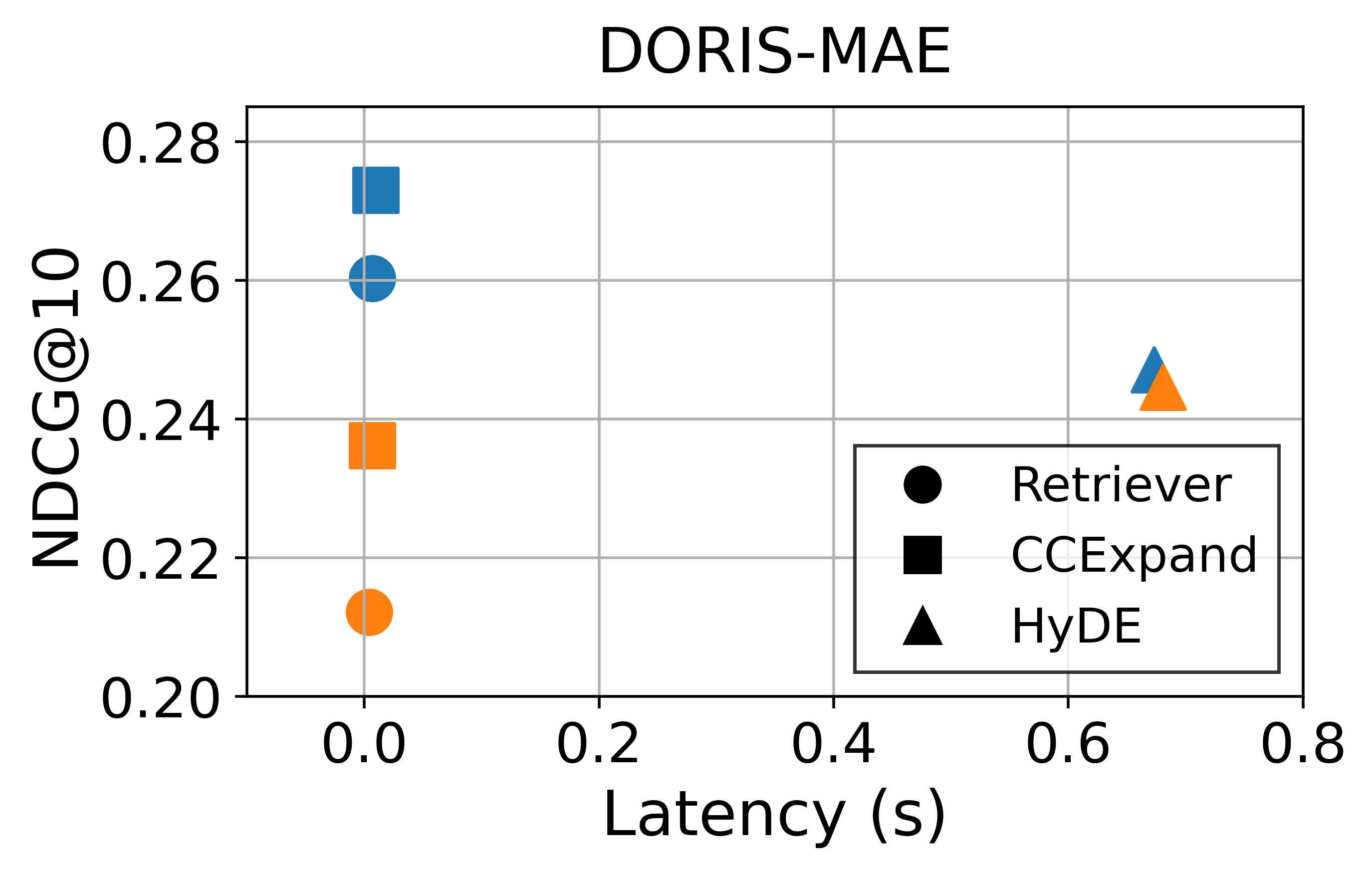}
        
        \label{fig:acc_latency_dorismae}
    \end{subfigure}
    
    \caption{Trade-off between retrieval effectiveness and efficiency. Blue and orange markers denote \ctr\ and \specter, respectively. We report the average time required to process a single query.}
    \label{fig:acc_latency}
\end{figure}


%% file: sections/99996_m_diversity.tex
\begin{figure*}[t]
    \centering
    \begin{subfigure}[b]{\textwidth}
        \centering
        \includegraphics[width=0.24\textwidth]{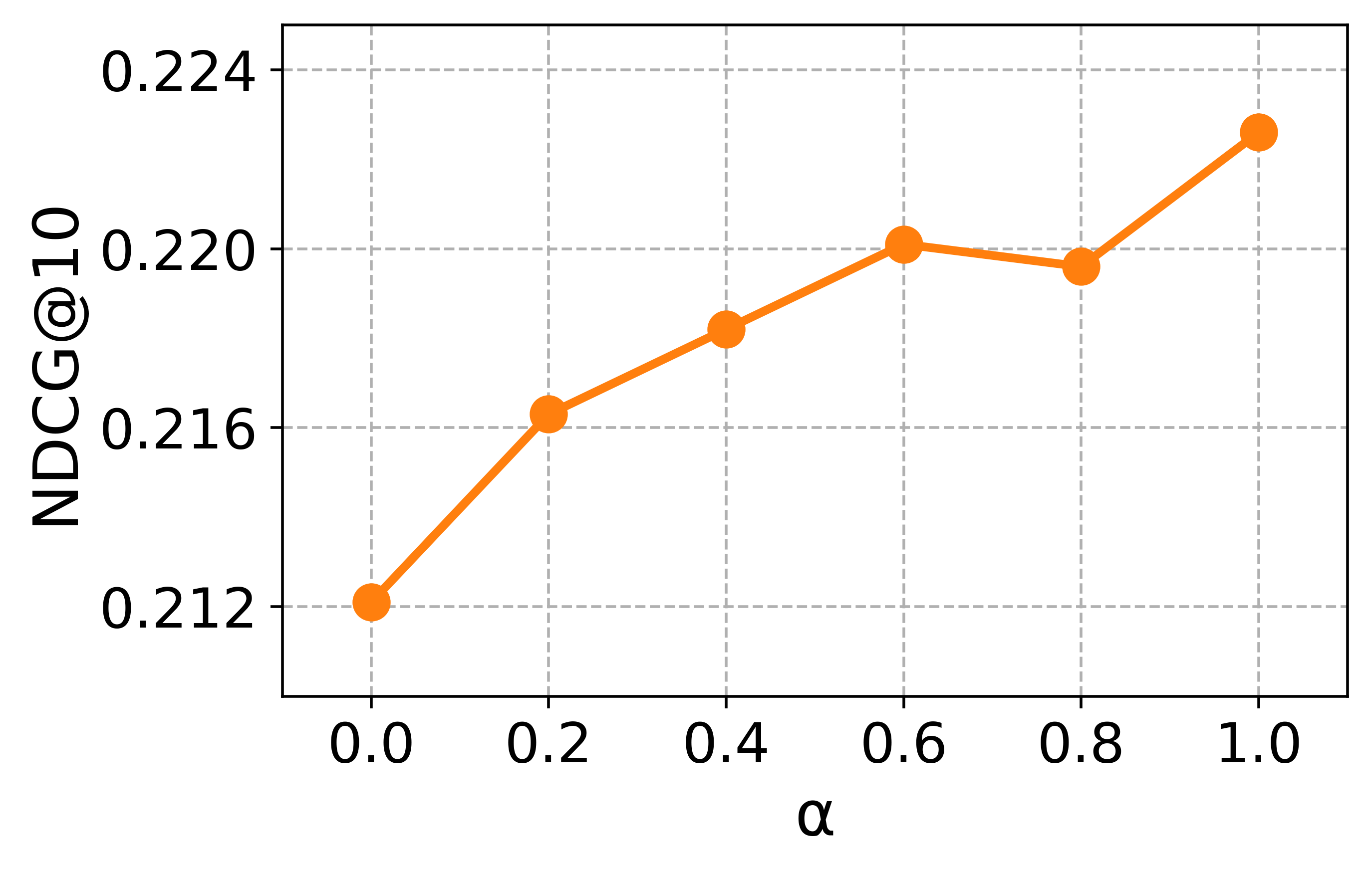}\hfill
        \includegraphics[width=0.24\textwidth]{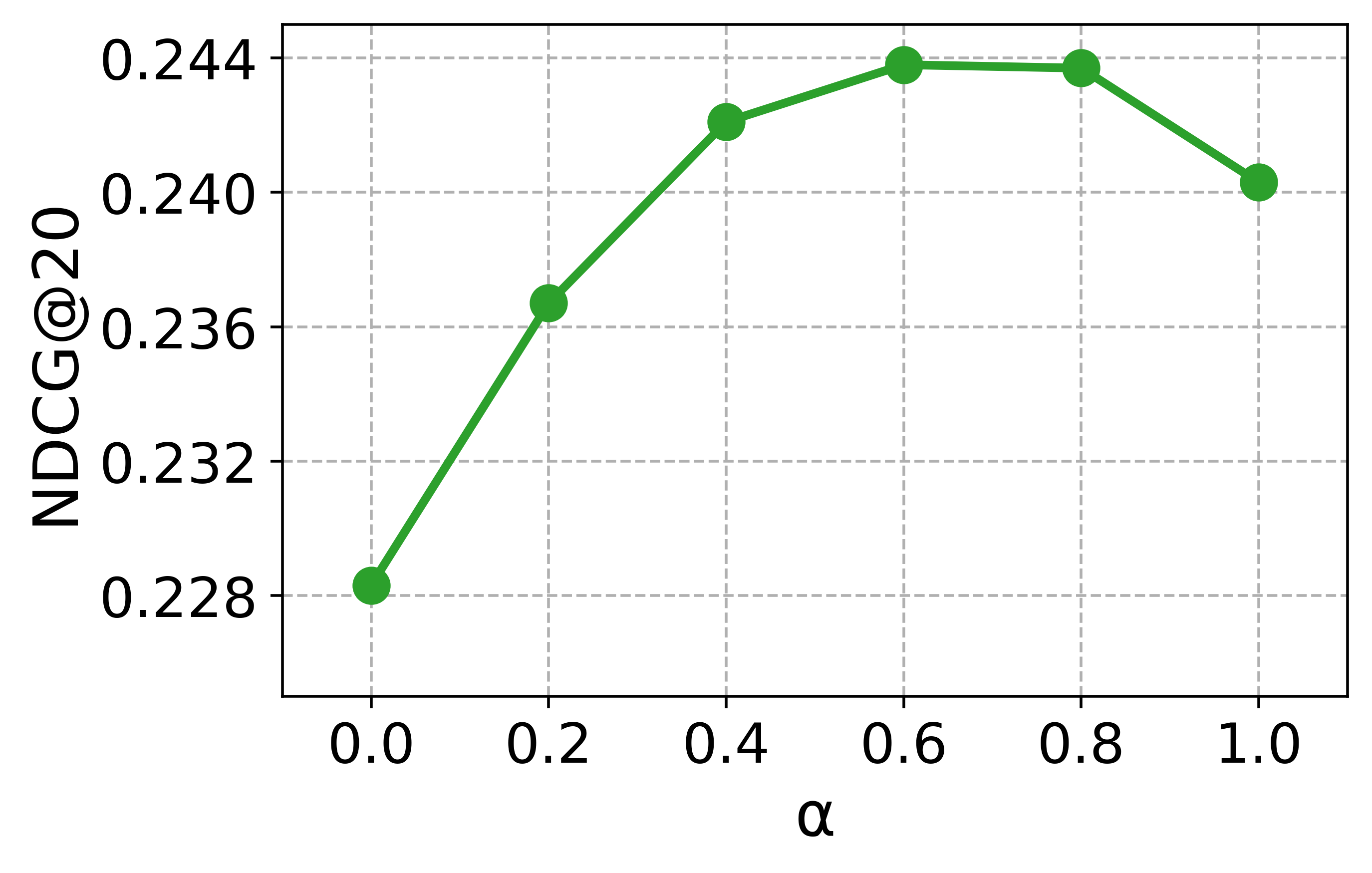}\hfill
        \includegraphics[width=0.24\textwidth]{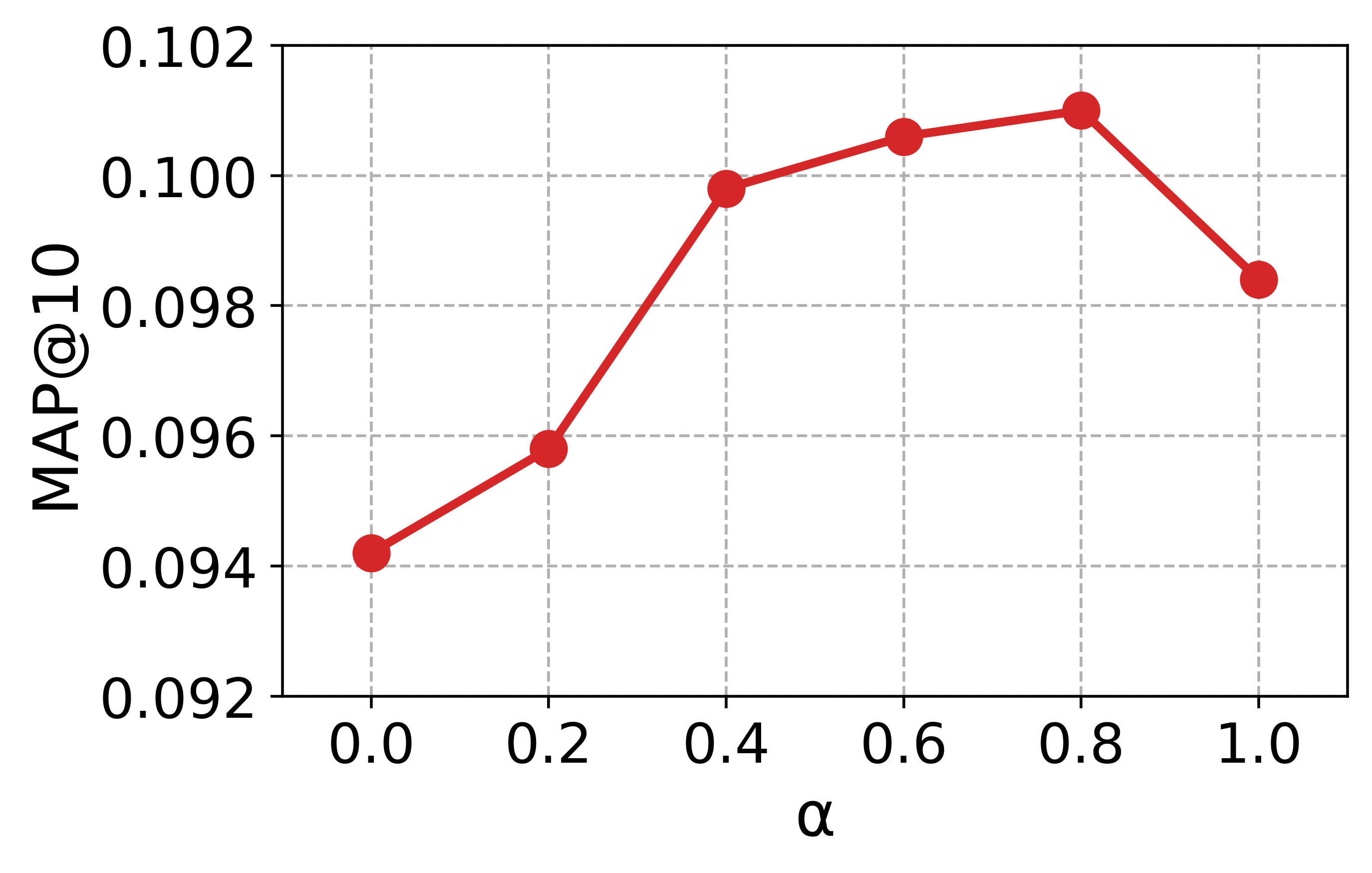}\hfill
        \includegraphics[width=0.24\textwidth]{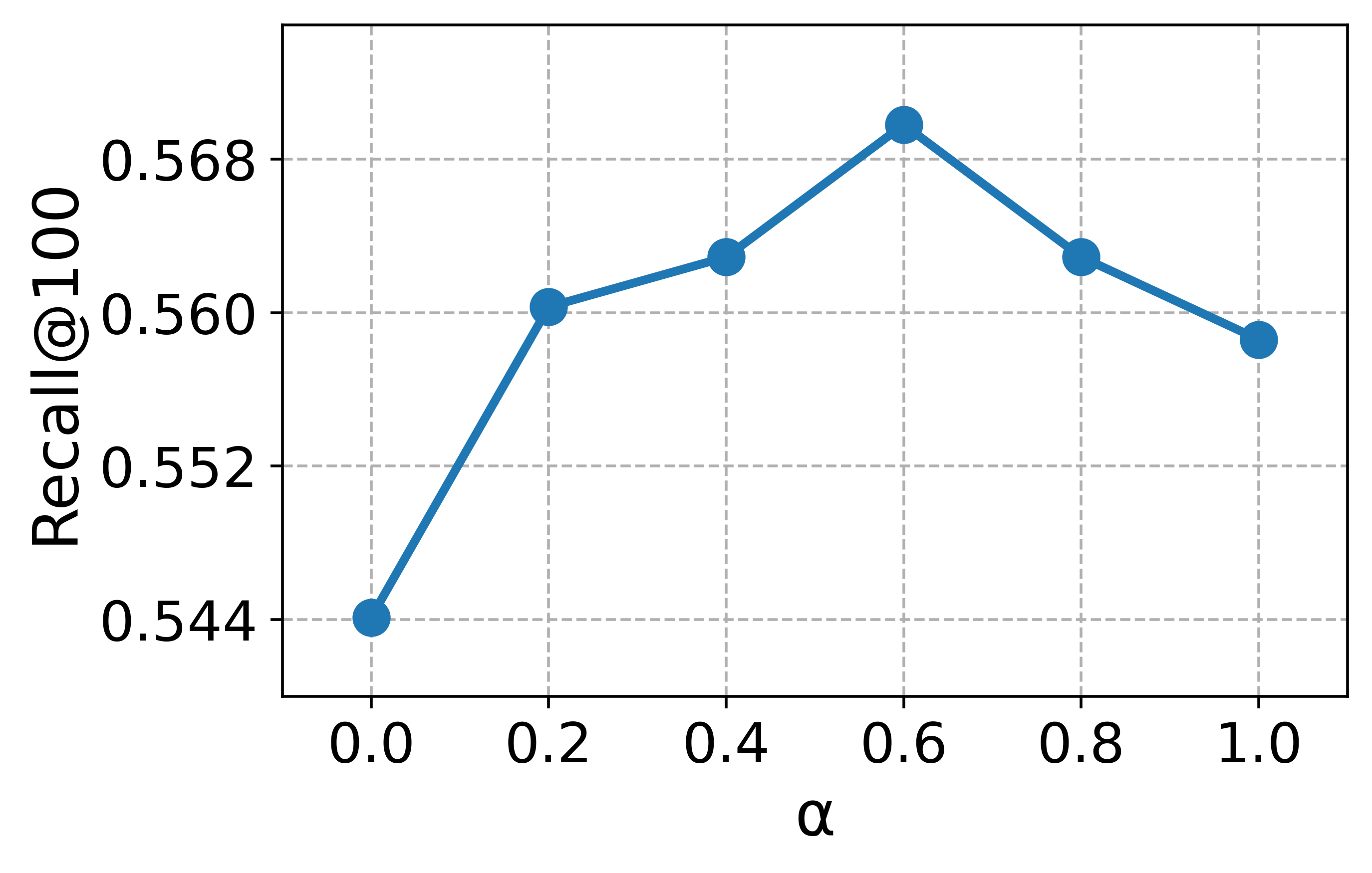}
        \caption{Impact of the balancing coefficient $\alpha$}
        \label{fig:alpha_analysis}
    \end{subfigure}
    
    \vspace{1em}

    \begin{subfigure}[b]{\textwidth}
        \centering
        \includegraphics[width=0.24\textwidth]{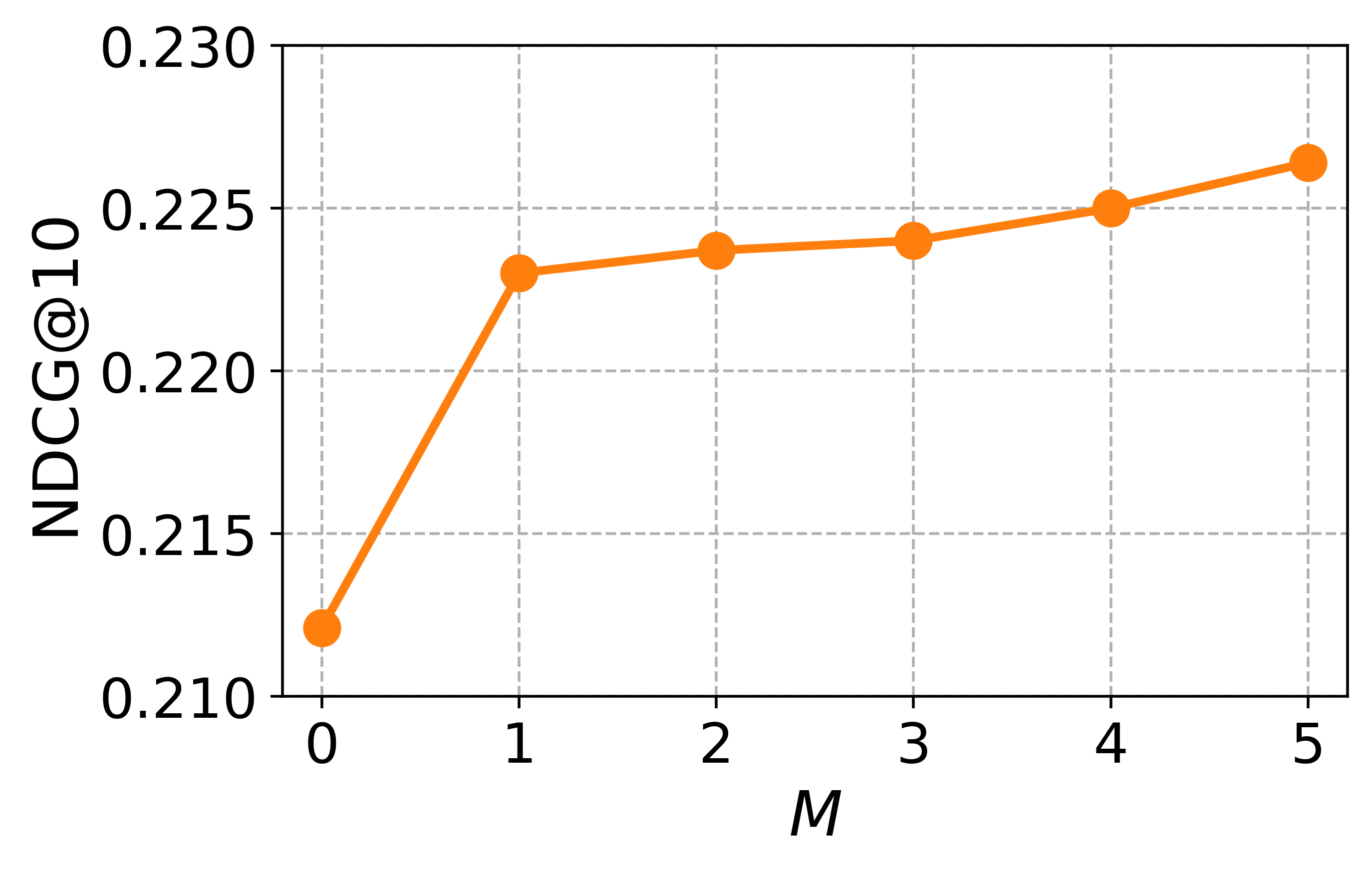}\hfill
        \includegraphics[width=0.24\textwidth]{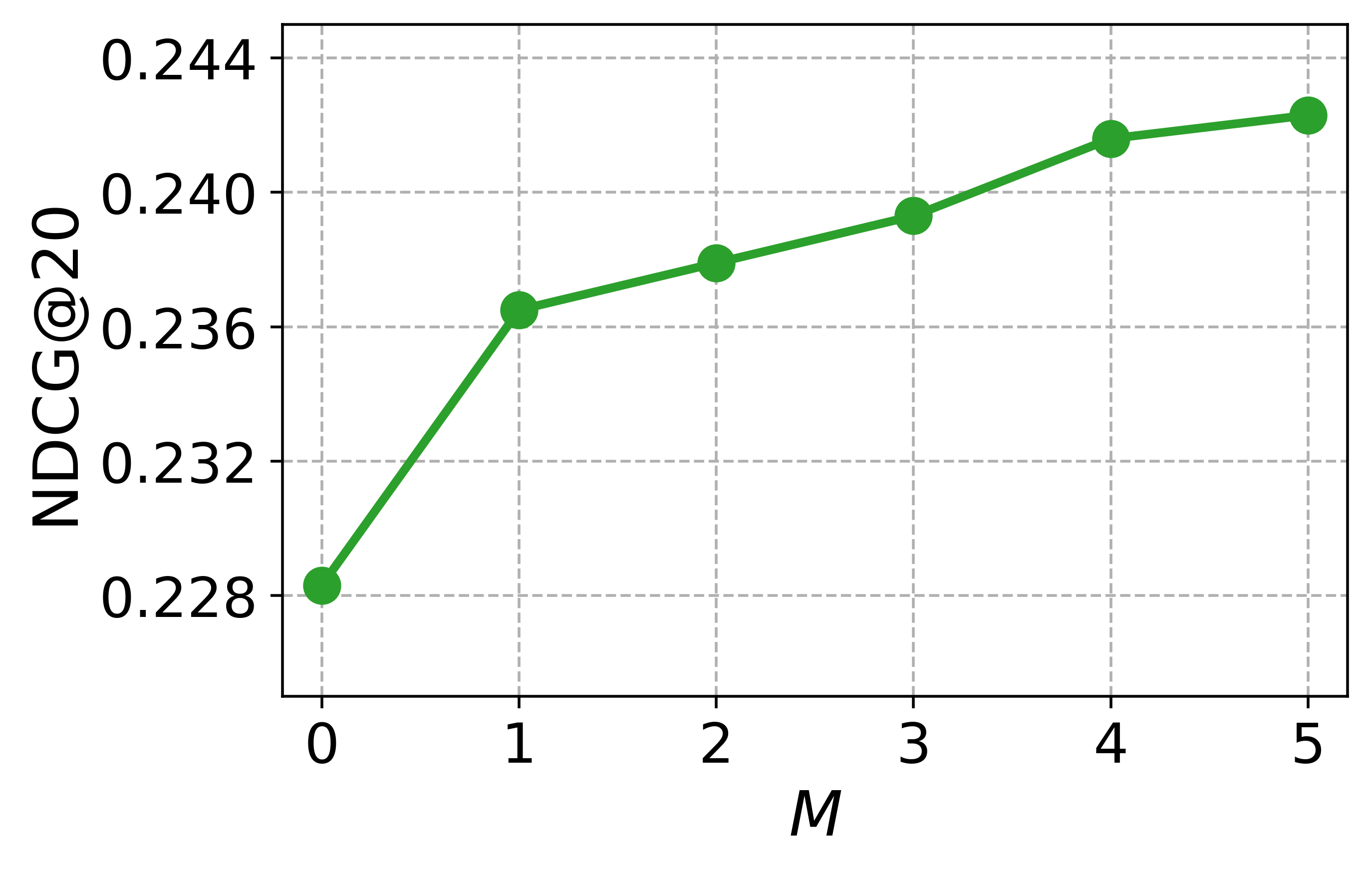}\hfill
        \includegraphics[width=0.24\textwidth]{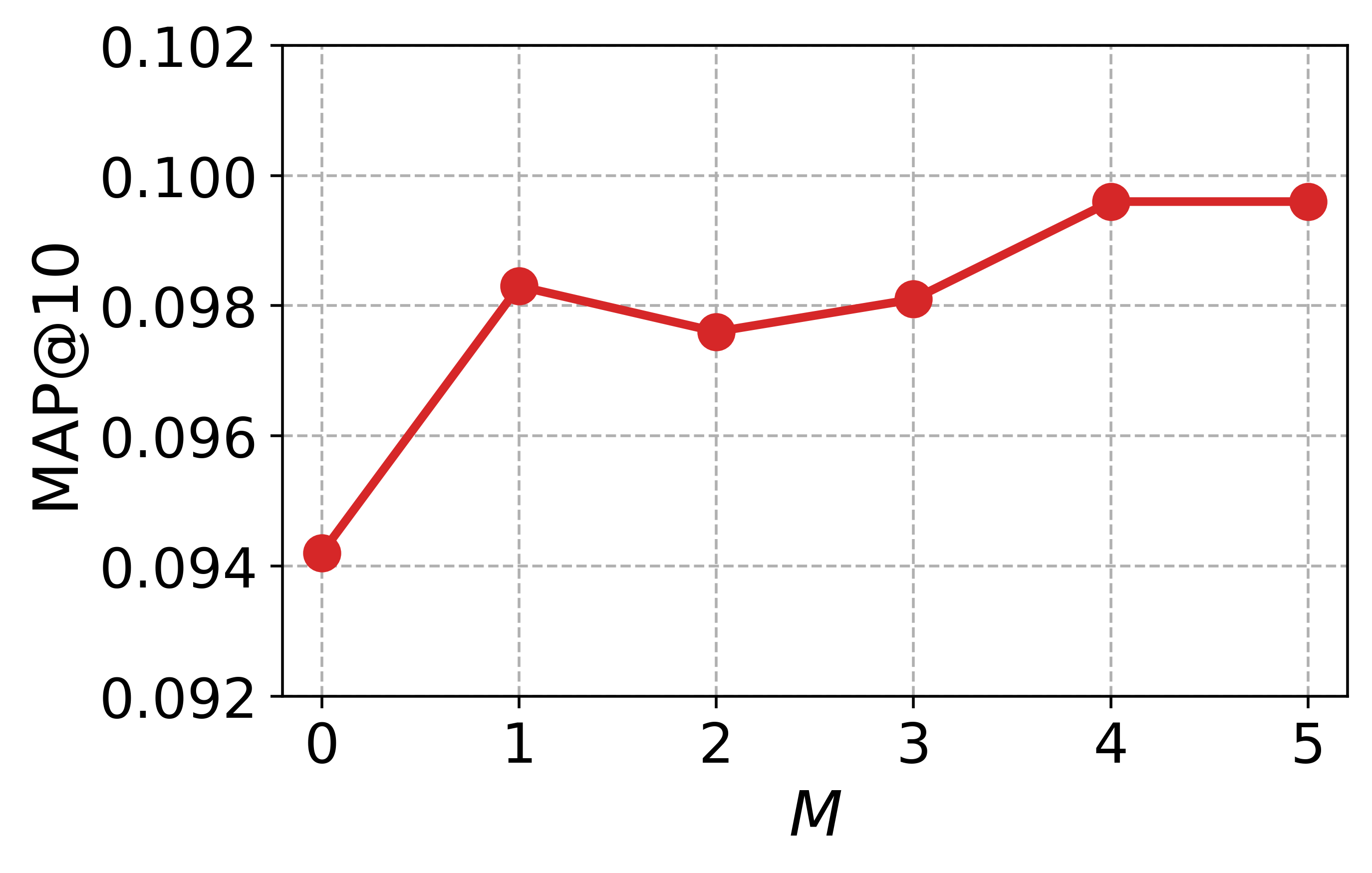}\hfill
        \includegraphics[width=0.24\textwidth]{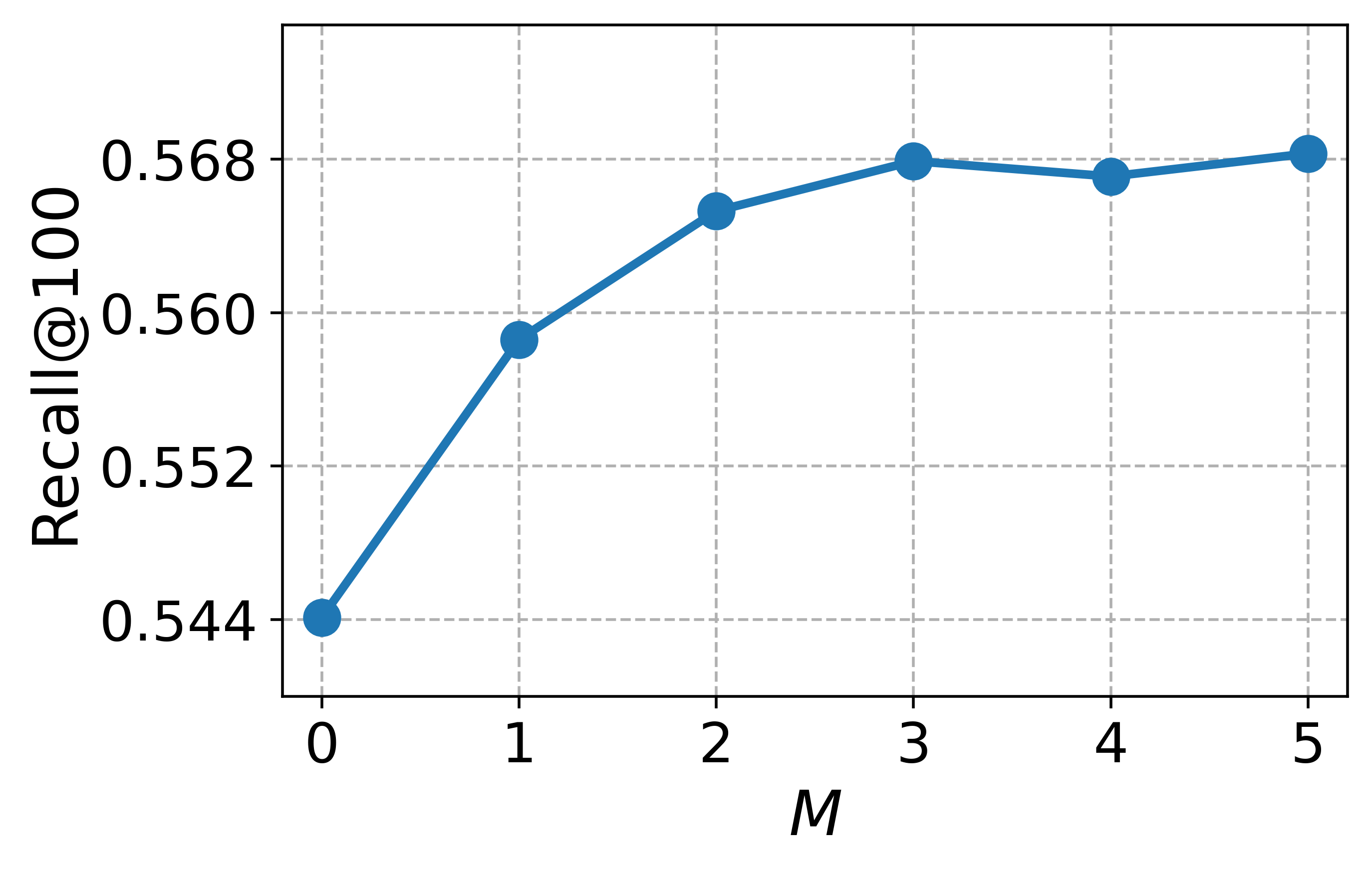}
        \caption{Impact of the number of snippets $M$}
        \label{fig:m_analysis}
    \end{subfigure}
    
    \caption{Hyperparameter analysis: (a) $\alpha$ (with fixed $M=5$), and (b) $M$ (with fixed $\alpha=0.6$). }
    \label{fig:parameter_anaylsis}
\end{figure*}

%% file: sections/99998_case.tex



\setlength{\fboxsep}{0pt}

\begin{table}[t]
\caption{\rev{Examples} of \proposedtwo. 
In both examples, the relevant document is initially ranked low (below 50).
By leveraging the most concept-aligned snippet during relevance matching, \proposedtwo retrieves the document at a substantially higher rank.
Contexts closely aligned with the query are highlighted in (\colorbox{softblue}{blue}).
}
\label{tab:rank_improvement_case}
\small
\centering
\resizebox{\linewidth}{!}{
\begin{tabular}[c]{p{1.1\linewidth}}
\toprule
\textbf{Case 1 (Rank : 70} $\rightarrow$ \textbf{2)} \\
\midrule
\textbf{User query} \\
I heard there is a technique that is called \colorbox{softblue}{behavioral cloning}, which allows the data to \colorbox{softblue}{learn from good players} initially. \\
\midrule
\textbf{Relevant document} \\
Imitating unknown policies via exploration. \colorbox{white}{Behavioral cloning} is an \colorbox{white}{imitation learning} technique that teaches an agent how to behave through \colorbox{white}{expert demonstrations}. ... We address these limitations incorporating a two-phase model into the original framework, which learns from unlabeled observations via exploration, substantially improving traditional \colorbox{white}{behavioral cloning} by exploiting a sampling mechanism to prevent bad local minima ... \\

\midrule

\textbf{Concept condition}: exploration, iterative, two phase model, self supervision\\

\textbf{Concept-aware query} \\
How does the integration of a two-phase model, utilizing \colorbox{softblue}{self-supervision for exploration}, address the issue of bad local minima in iterative learning processes, as discussed in the paper? \\

\textbf{Concept-focused snippet} \\
The document presents a novel approach to \colorbox{softblue}{behavioral cloning} that incorporates a two-phase model designed to tackle the problem of getting trapped in bad local minima during iterative learning processes. 
By integrating \colorbox{softblue}{self-supervision and a robust exploration} strategy, the model enhances the \colorbox{softblue}{learning from unlabeled observations} ... demonstrating an effective solution to the challenges posed by local minima in \colorbox{softblue}{imitation learning}. \\

\midrule
\textbf{Case 2 (Rank : 59} $\rightarrow$ \textbf{5)} \\
\midrule
\textbf{User query} \\
I want to improve the \colorbox{softblue}{interpretability and explainability of machine learning models} in various fields such as healthcare and finance by using \colorbox{softblue}{representation learning}. \\
\midrule

\textbf{Relevant document } \\

An analysis on the use of autoencoders for \colorbox{white}{representation learning}: fundamentals, learning task case studies, \colorbox{white}{explainability} and challenges. In many machine learning tasks, learning a good representation of the data can be the key to building a well-performant solution. ... The theoretical developments are put into practice using a selection of datasets for the different problems and implementing each solution, ... We also explore the current challenges and approaches to \colorbox{white}{explainability} in the context of autoencoders. All of this helps conclude that, thanks to alterations in their structure as well as their objective function, autoencoders may be the core of a possible solution to many problems ...
\\
\midrule
\textbf{Concept condition}: \colorbox{white}{representation learning}, autoencoder, \colorbox{white}{explainability}\\
\textbf{Concept-aware query} \\
How do autoencoders contribute to the \colorbox{softblue}{explainability} of models in the context of \colorbox{softblue}{representation learning}, and what are the current methodologies for \colorbox{softblue}{interpreting and understanding} the learned representations generated by autoencoders? \\

\textbf{Concept-focused snippet} \\
The document addresses the query by exploring the role of autoencoders in enhancing the \colorbox{softblue}{explainability of machine learning models} within the domain of \colorbox{softblue}{representation learning}. 
It discusses current challenges related to understanding and \colorbox{softblue}{interpreting the learned representations} generated by autoencoders, highlighting methodologies that can be employed to \colorbox{softblue}{improve interpretability}. The paper emphasizes that modifications in their architecture or objective functions can lead to more intuitive and \colorbox{softblue}{interpretable results}. ... \\
\bottomrule

\end{tabular}}
\end{table}

%% file: sections/070Conclusion.tex
We introduce the academic concept index, a structured representation of document-level concept information, and investigate how it can systematically guide two major lines of retrieval enhancement.
First, we propose \proposed, a concept-aware synthetic query generation method that adaptively conditions LLMs on uncovered concepts to produce complementary queries with broader conceptual coverage. 
Second, we introduce \proposedtwo, a training-free context-augmentation method that generates concept-focused snippets and strengthens relevance matching through concept-centric signals. 
These two components provide complementary benefits: \proposed improves the quality and diversity of training data for fine-tuning, whereas \proposedtwo enhances inference-time relevance estimation without requiring any retriever updates.
Extensive experiments validate that the proposed methods substantially improve conceptual alignment, reduce redundancy, and boost retrieval performance.
We acknowledge that the proposed methods involve multiple stages (e.g., concept extraction, query generation, and snippet generation), which may introduce additional complexity compared to simpler baselines.
However, the pipeline is modular and fully offline, incurring no additional cost at inference time.
Simplifying this pipeline while preserving its effectiveness remains an important direction for future work.
Beyond retrieval, we also plan to explore the broader utility of the academic concept index in other tasks such as paper-reviewer matching~\cite{zhang2025chain}.


%% file: sections/080Appendix.tex
\section{\rev{Dataset Statistics}}
\label{appendix:data_statistics}
\input{sections/99997_taxonomy_ablation}
\rev{Table~\ref{tab:data_statistics} summarizes the key statistics of both datasets.
\csfcube dataset consists of 50 test queries, with about 120 candidates per query drawn from approximately 800,000 papers in the S2ORC corpus \cite{lo2020s2orc}. 
\dorismae dataset consists of 165,144 test queries, with candidates drawn similarly to \csfcube.
The two datasets also differ substantially in annotation methodology: \csfcube provides a smaller set of expert-annotated query-by-example pairs with dense relevance judgments per query, while \dorismae offers a much larger collection of human-written queries with LLM-based annotations, resulting in sparser but broader coverage.
We consider annotation scores above `2', which indicate documents are `nearly identical or similar' (\csfcube) and `directly answer all key components' (\dorismae), as relevant.
}

\begin{table}[h]
\centering
\caption{\rev{Dataset statistics for \csfcube and \dorismae.}}
\label{tab:data_statistics}
\revcolor
\begin{tabular}{lrr}
\hline
\textbf{Statistic} & \textbf{\csfcube} & \textbf{\dorismae} \\
\hline
\# Test queries & 50 &  165,144\\
\# Candidate documents & 4,207 & 8,482 \\
Avg. relevant docs / query & 13.32 & 10.84 \\
Annotation method & Human experts & LLMs \\
\hline
\end{tabular}
\end{table}

\section{\rev{Additional Examples}}
\label{appendix:examples}
\rev{We provide additional examples of the queries from \proposed and the snippets from \proposedtwo.}

\rev{Table~\ref{tab:case_study} compares synthetic queries generated with and without \proposed, using Promptgator as the default prompting method. 
Across both examples, queries generated by \proposed exhibit enhanced term usage and broader concept coverage, explicitly targeting concepts such as \textit{implicit differentiation} and \textit{mel-spectrogram features} that are underrepresented in the baseline queries. 
This supports the quantitative results in Section~\ref{sec:experimentresult}, showing that \proposed produces more complementary and conceptually diverse training queries.}

\rev{
Table~\ref{tab:bad_case} presents a challenging example of \proposed on a game theory paper from DORIS-MAE. 
When the academic taxonomy provides limited coverage for certain domains (e.g., game theory, social science), the concept extractor may place relatively higher weight on generic terms (e.g., \textit{implementation}, \textit{session}, \textit{process}) than on domain-specific terminology (e.g., \textit{reflexive game theory}, \textit{group influence}), which can bias the generated queries toward more general aspects of the document.
This observation highlights the importance of ensuring sufficient topic coverage, particularly for interdisciplinary domains.
Even when the generated queries deviate from the core domain focus, \proposedtwo's concept-focused snippet remains grounded in the original document, as it directly conditions on the document content rather than relying solely on the query. As a result, it preserves domain-relevant information and mitigates the impact of such query-level deviations.
}

\input{sections/9999CCQGen_casestudy}
\input{sections/9999d_CCQGen_badcase}

%% file: sections/9999CCQGen_casestudy.tex
\begin{table}[t]
\caption{\rev{Examples} comparing synthetic queries generated without and with \proposed.
Each example follows a consistent structure: (1) the document, (2) queries generated without CCQGen \rev{(i.e., Promptgator)}, and (3) queries generated with CCQGen \rev{(i.e., Promptgator w/ CCQGen)}.}
\small
\centering
\resizebox{\linewidth}{!}{
\begin{tabular}{C}
\toprule
\rev{\textbf{Example 1}} \\
\midrule
\textbf{Document (truncated)} \\
A core capability of intelligent systems is the ability to quickly learn new tasks by drawing on prior experience. 
Gradient (or optimization) based meta-learning has recently emerged as an effective approach for few-shot learning. ...
By drawing upon implicit differentiation, we develop the implicit MAML algorithm, which depends only on the solution to the inner level optimization and not the path taken by the inner loop optimizer. ...
Theoretically, we prove that implicit MAML can compute accurate meta-gradients with a memory footprint ... \\
\midrule
\textbf{Queries without CCQGen} \\
\begin{enumerate}[leftmargin=*]
    \item What are the advantages of implicit MAML compared to traditional gradient-based meta-learning approaches?
    \item In what ways does the implicit MAML algorithm demonstrate empirical gains on few-shot image recognition benchmarks?
    \item How does the implicit MAML algorithm contribute to the scalability of gradient-based meta-learning approaches?
\end{enumerate} \\
\midrule
\textbf{Queries with CCQGen} \\
\begin{enumerate}[leftmargin=*]
    \item How does implicit differentiation improve efficiency in meta-learning within the implicit MAML framework?
    \item How does implicit MAML support accurate meta-parameter computation when scaling meta-learning systems?
    \item How do few-shot learning scenarios benefit from efficient meta-gradient computation enabled by implicit MAML?
\end{enumerate} \\
\midrule
\rev{\textbf{Example 2}} \\
\midrule
\textbf{Document} \\
Automated music playlist generation is a specific form of music recommendation. ...
However, the scarcity of thoroughly curated playlists and the bias towards popular songs result in the vast majority of songs occurring in very few playlists and thus being poorly recommended.
To overcome this issue, we propose an alternative model based on a song-to-playlist classifier, which learns the underlying structure from actual playlists while leveraging song features derived from audio, social tags and independent listening logs. ...
\\
\midrule
\textbf{Queries without CCQGen} \\
\begin{enumerate}[leftmargin=*]
    \item How can song-to-playlist classifiers enhance automated music playlist generation?
    \item How can automated playlist creation be improved through song-to-playlist classification and feature exploitation?
    \item How does a song-to-playlist classifier differ from traditional collaborative filtering for music recommendation?
\end{enumerate} \\
\midrule
\textbf{Queries with CCQGen} \\
\begin{enumerate}[leftmargin=*]
    \item What techniques can be used to overcome filter bubbles and improve recommendation of out-of-set songs?
    \item How can mel-spectrogram features help mitigate the cold-start problem in playlist recommendation?
    \item How can audio-based representations improve recommendation accuracy for rare or under-represented songs?
\end{enumerate} \\
\bottomrule
\end{tabular}}
\label{tab:case_study}
\end{table}

%% file: sections/9999d_CCQGen_badcase.tex
\begin{table}[t]
\caption{\rev{
An illustrative challenging example for CCQGen on a game theory paper (DORIS-MAE). 
Query $q_2$ partially deviates from the core game-theoretic context, likely due to limited taxonomy coverage in interdisciplinary domains (e.g., game theory and social science).
Nevertheless, CCExpand remains effective by preserving relevant content through grounding in the original document.
}}
\small
\centering
\resizebox{\linewidth}{!}{
\begin{tabular}{lp{0.38\linewidth}p{0.53\linewidth}}
\toprule

\multicolumn{3}{l}{\textbf{Document}} \\
\midrule
\multicolumn{3}{p{0.95\linewidth}}{
Modeling multistage decision processes with Reflexive Game Theory. This paper introduces application of Reflexive Game Theory to the matter of multistage decision making processes. The idea behind is that each decision making session has certain parameters like ``who are the group members'', ``how group members influence on each other'', etc. This study illustrates the consecutive or sequential  \ldots
} \\

\midrule

 & \textbf{Keywords} & \textbf{CCQGen Query} \\
\midrule

$q_1$ & group members, influence, process, decision
  & How does the model handle the influence of group members on each other during the decision making process? \\[4pt]

\textcolor{red}{$q_2$} & parameter, final, implementation, session
  & \textcolor{red}{How are parameters determined for the final implementation session in the multistage decision-making process?} \\[4pt]

$q_3$ & decision making, ultimate, implementation, application
  & How is the Reflexive Game Theory applied to the ultimate decision-making process during implementation stages? \\[4pt] 

\midrule

\multicolumn{3}{l}{\textbf{\proposedtwo Snippet (conditioned on $q_2$)}} \\
\midrule
\multicolumn{3}{p{0.95\linewidth}}{
In the multistage decision-making process, the parameters of the final implementation session are determined during earlier decision-making stages. Specifically, stage 1 focuses on deciding key aspects such as the timing of the session, the participating group members, and the influence structure among them. This stage may involve multiple intermediate decision sessions to refine these parameters. Once established, stage 2 proceeds with the implementation of the final decision based on the configured parameters.
} \\

\bottomrule
\end{tabular}}
\label{tab:bad_case}
\end{table}

%% file: main.bib
@inproceedings{SPECTER,
  title={{SPECTER: Document-level Representation Learning using Citation-informed Transformers}},
  author={Arman Cohan and Sergey Feldman and Iz Beltagy and Doug Downey and Daniel S. Weld},
  booktitle={ACL},
  year={2020}
}

@inproceedings{MAG_FS,
  title={A Web-scale system for scientific knowledge exploration},
  author={Shen, Zhihong and Ma, Hao and Wang, Kuansan},
  booktitle={ACL},
  pages={87--92},
  year={2018}
}

@inproceedings{lee2022taxocom,
  title={Taxocom: Topic taxonomy completion with hierarchical discovery of novel topic clusters},
  author={Lee, Dongha and Shen, Jiaming and Kang, SeongKu and Yoon, Susik and Han, Jiawei and Yu, Hwanjo},
  booktitle={WWW},
  pages={2819--2829},
  year={2022}
}

@article{autophrase,
  title={Automated phrase mining from massive text corpora},
  author={Shang, Jingbo and Liu, Jialu and Jiang, Meng and Ren, Xiang and Voss, Clare R and Han, Jiawei},
  journal={IEEE Transactions on Knowledge and Data Engineering},
  volume={30},
  number={10},
  pages={1825--1837},
  year={2018},
  publisher={IEEE}
}

@inproceedings{mmoe,
  title={Modeling task relationships in multi-task learning with multi-gate mixture-of-experts},
  author={Ma, Jiaqi and Zhao, Zhe and Yi, Xinyang and Chen, Jilin and Hong, Lichan and Chi, Ed H},
  booktitle={KDD},
  pages={1930--1939},
  year={2018}
}

@inproceedings{
    BEIR,
    title={{BEIR}: A Heterogeneous Benchmark for Zero-shot Evaluation of Information Retrieval Models},
    author={Nandan Thakur and Nils Reimers and Andreas R{\"u}ckl{\'e} and Abhishek Srivastava and Iryna Gurevych},
    booktitle={NeurIPS Datasets and Benchmarks Track},
    year={2021},
}

@inproceedings{dai2022promptagator,
  title={Promptagator: Few-shot dense retrieval from 8 examples},
  author={Dai, Zhuyun and Zhao, Vincent Y and Ma, Ji and Luan, Yi and Ni, Jianmo and Lu, Jing and Bakalov, Anton and Guu, Kelvin and Hall, Keith B and Chang, Ming-Wei},
  booktitle={ICLR},
  year={2023}
}

@inproceedings{DPR,
    title = "Dense Passage Retrieval for Open-Domain Question Answering",
    author = "Karpukhin, Vladimir and Oguz, Barlas and Min, Sewon and Lewis, Patrick and Wu, Ledell and Edunov, Sergey and Chen, Danqi and Yih, Wen-tau",
    booktitle = "EMNLP",
    year = "2020",
    pages = "6769--6781",
}

@inproceedings{zhan2021optimizing,
  title={Optimizing dense retrieval model training with hard negatives},
  author={Zhan, Jingtao and Mao, Jiaxin and Liu, Yiqun and Guo, Jiafeng and Zhang, Min and Ma, Shaoping},
  booktitle={SIGIR},
  pages={1503--1512},
  year={2021}
}

@inproceedings{rocketqa_v1,
    title = "{R}ocket{QA}: An Optimized Training Approach to Dense Passage Retrieval for Open-Domain Question Answering",
    author = "Qu, Yingqi  and
      Ding, Yuchen  and
      Liu, Jing  and
      Liu, Kai  and
      Ren, Ruiyang  and
      Zhao, Wayne Xin  and
      Dong, Daxiang  and
      Wu, Hua  and
      Wang, Haifeng",
    booktitle = "NAACL-HLT",
    year = "2021",
    pages = "5835--5847"
}

@article{CTR,
  title={Unsupervised dense information retrieval with contrastive learning},
  author={Izacard, Gautier and Caron, Mathilde and Hosseini, Lucas and Riedel, Sebastian and Bojanowski, Piotr and Joulin, Armand and Grave, Edouard},
  journal={arXiv preprint arXiv:2112.09118},
  year={2021}
}

@inproceedings{AR2,
  title={Adversarial Retriever-Ranker model for Dense Retrieval},
  author={Zhang, Hang and Gong, Yeyun and Shen, Yelong and Lv, Jiancheng and Duan, Nan and Chen, Weizhu},
  booktitle={ICLR},
  year={2022}
}

@inproceedings{SCIBERT,
  title={SciBERT: Pretrained Language Model for Scientific Text},
  author={Iz Beltagy and Kyle Lo and Arman Cohan},
  year={2019},
  booktitle={EMNLP},
  pages = "3615--3620"
}

@inproceedings{SCINCL,
  author = {Malte Ostendorff and Nils Rethmeier and Isabelle Augenstein and Bela Gipp and Georg Rehm},
  title = {{Neighborhood Contrastive Learning for Scientific Document Representations with Citation Embeddings}},
  booktitle = {EMNLP},
  year = 2022
}

@inproceedings{OAGBERT,
author = {Liu, Xiao and Yin, Da and Zheng, Jingnan and Zhang, Xingjian and Zhang, Peng and Yang, Hongxia and Dong, Yuxiao and Tang, Jie},
title = {OAG-BERT: Towards a Unified Backbone Language Model for Academic Knowledge Services},
year = {2022},
booktitle = {KDD},
pages = {3418–3428}
}

@inproceedings{ASPIRE,
  title={Multi-Vector Models with Textual Guidance for Fine-Grained Scientific Document Similarity},
  author={Mysore, Sheshera and Cohan, Arman and Hope, Tom},
  booktitle={NAACL},
  pages={4453--4470},
  year={2022}
}

@inproceedings{SPECTER2,
  title={SciRepEval: A Multi-Format Benchmark for Scientific Document Representations},
  author={Singh, Amanpreet and D’Arcy, Mike and Cohan, Arman and Downey, Doug and Feldman, Sergey},
  booktitle={EMNLP},
  pages={5548--5566},
  year={2023}
}

@inproceedings{zhang2023pre,
  title={Pre-training Multi-task Contrastive Learning Models for Scientific Literature Understanding},
  author={Zhang, Yu and Cheng, Hao and Shen, Zhihong and Liu, Xiaodong and Wang, Ye-Yi and Gao, Jianfeng},
  booktitle={Findings of EMNLP},
  pages={12259--12275},
  year={2023}
}

@inproceedings{mackie2023generative,
  title={Generative Relevance Feedback with Large Language Models},
  author={Mackie, Iain and Chatterjee, Shubham and Dalton, Jeffrey},
  booktitle={SIGIR},
pages = {2026–2031},
  year={2023}
}

@inproceedings{ToTER,
  title={Improving Retrieval in Theme-specific Applications using a Corpus Topical Taxonomy},
  author={Kang, SeongKu and Agarwal, Shivam and Jin, Bowen and Lee, Dongha and Yu, Hwanjo and Han, Jiawei},
  booktitle={WWW},
  pages = {1497–1508},
  year={2024}
}

@inproceedings{mao2021generation,
  title={Generation-Augmented Retrieval for Open-Domain Question Answering},
  author={Mao, Yuning and He, Pengcheng and Liu, Xiaodong and Shen, Yelong and Gao, Jianfeng and Han, Jiawei and Chen, Weizhu},
  booktitle={ACL},
  pages={4089--4100},
  year={2021}
}

@inproceedings{li2023sailer,
  title={SAILER: Structure-aware Pre-trained Language Model for Legal Case Retrieval},
  author={Li, Haitao and Ai, Qingyao and Chen, Jia and Dong, Qian and Wu, Yueyue and Liu, Yiqun and Chen, Chong and Tian, Qi},
    booktitle={SIGIR},
  year={2023}
}

@article{CSFCube,
  title={CSFCube-A Test Collection of Computer Science Research Articles for Faceted Query by Example},
  author={Mysore, Sheshera and O'Gorman, Tim and McCallum, Andrew and Zamani, Hamed},
  journal={NeurIPS 2021 Track on Datasets and Benchmarks},
  year={2021}
}

@inproceedings{
DORISMAE,
title={Scientific Document Retrieval using Multi-level Aspect-based Queries},
author={Jianyou Wang and Kaicheng Wang and Xiaoyue Wang and Prudhviraj Naidu and Leon Bergen and Ramamohan Paturi},
booktitle={NeurIPS Datasets and Benchmarks Track},
year={2023}
}

@article{tao2016multi,
  title={Multi-Dimensional, Phrase-Based Summarization in Text Cubes},
  author={Tao, Fangbo and Zhuang, Honglei and Yu, Chi Wang and Wang, Qi and Cassidy, Taylor and Kaplan, Lance M and Voss, Clare R and Han, Jiawei},
  journal={IEEE Data Eng. Bull.},
  volume={39},
  number={3},
  pages={74--84},
  year={2016}
}

@inproceedings{formal2022distillation,
  title={From distillation to hard negative sampling: Making sparse neural ir models more effective},
  author={Formal, Thibault and Lassance, Carlos and Piwowarski, Benjamin and Clinchant, St{\'e}phane},
  booktitle={SIGIR},
  pages={2353--2359},
  year={2022}
}

@inproceedings{shi2024taxonomy,
  title={Taxonomy completion via implicit concept insertion},
  author={Shi, Jingchuan and Dong, Hang and Chen, Jiaoyan and Wu, Zhe and Horrocks, Ian},
  booktitle={WWW},
  pages={2159--2169},
  year={2024}
}

@inproceedings{inpars,
  title={Inpars: Unsupervised dataset generation for information retrieval},
  author={Bonifacio, Luiz and Abonizio, Hugo and Fadaee, Marzieh and Nogueira, Rodrigo},
  booktitle={SIGIR},
  pages={2387--2392},
  year={2022}
}

@inproceedings{pairwise_qgen,
  title={It's All Relative!--A Synthetic Query Generation Approach for Improving Zero-Shot Relevance Prediction},
  author={Chaudhary, Aditi and Raman, Karthik and Bendersky, Michael},
  booktitle={Findings of NAACL},
  year={2024}
}

@inproceedings{tong2025igft,
  title={From Missteps to Mastery: Enhancing Low-Resource Dense Retrieval through Adaptive Query Generation},
  author={Tong, Zhenyu and Qin, Chuan and Fang, Chuyu and Yao, Kaichun and Chen, Xi and Zhang, Jingshuai and Zhu, Chen and Zhu, Hengshu},
  booktitle={Proceedings of the 31st ACM SIGKDD Conference on Knowledge Discovery and Data Mining V.1 (KDD '25)},
  pages={1373--1384},
  year={2025}
}

@inproceedings{kim2025syntriever,
  title={Syntriever: How to Train Your Retriever with Synthetic Data from LLMs},
  author={Kim, Minsang and Baek, Seung Jun},
  booktitle={Findings of the Association for Computational Linguistics: NAACL 2025},
  pages={2523--2539},
  year={2025}
}

@inproceedings{label_condition_qgen,
  title={Exploring the viability of synthetic query generation for relevance prediction},
  author={Chaudhary, Aditi and Raman, Karthik and Srinivasan, Krishna and Hashimoto, Kazuma and Bendersky, Mike and Najork, Marc},
  booktitle={The SIGIR 2023 Workshop on eCommerce},
  year={2023}
}

@article{inpars2,
  title={Inpars-v2: Large language models as efficient dataset generators for information retrieval},
  author={Jeronymo, Vitor and Bonifacio, Luiz and Abonizio, Hugo and Fadaee, Marzieh and Lotufo, Roberto and Zavrel, Jakub and Nogueira, Rodrigo},
  journal={arXiv preprint arXiv:2301.01820},
  year={2023}
}

@inproceedings{control_gen,
  title={Controlled text generation with natural language instructions},
  author={Zhou, Wangchunshu and Jiang, Yuchen Eleanor and Wilcox, Ethan and Cotterell, Ryan and Sachan, Mrinmaya},
  booktitle={ICML},
  pages={42602--42613},
  year={2023},
}

@inproceedings{outline_condition,
  title={Advancing Precise Outline-Conditioned Text Generation with Task Duality and Explicit Outline Control},
  author={Li, Yunzhe and Chen, Qian and Yan, Weixiang and Wang, Wen and Zhang, Qinglin and Sundaram, Hari},
  booktitle={EACL},
  pages={2362--2377},
  year={2024}
}

@inproceedings{alberti2019synthetic,
  title={Synthetic QA Corpora Generation with Roundtrip Consistency},
  author={Alberti, Chris and Andor, Daniel and Pitler, Emily and Devlin, Jacob and Collins, Michael},
  booktitle={ACL},
  pages={6168--6173},
  year={2019}
}

@inproceedings{sachan2022improving,
  title={Improving Passage Retrieval with Zero-Shot Question Generation},
  author={Sachan, Devendra and Lewis, Mike and Joshi, Mandar and Aghajanyan, Armen and Yih, Wen-tau and Pineau, Joelle and Zettlemoyer, Luke},
  booktitle={EMNLP},
  pages={3781--3797},
  year={2022}
}

@inproceedings{saad2023udapdr,
  title={UDAPDR: Unsupervised Domain Adaptation via LLM Prompting and Distillation of Rerankers},
  author={Saad-Falcon, Jon and Khattab, Omar and Santhanam, Keshav and Florian, Radu and Franz, Martin and Roukos, Salim and Sil, Avirup and Sultan, Md and Potts, Christopher},
  booktitle={EMNLP},
  pages={11265--11279},
  year={2023}
}

@inproceedings{GPT3,
  title={Language models are few-shot learners},
  author={Brown, Tom and Mann, Benjamin and Ryder, Nick and Subbiah, Melanie and Kaplan, Jared D and Dhariwal, Prafulla and Neelakantan, Arvind and Shyam, Pranav and Sastry, Girish and Askell, Amanda and others},
  booktitle={NeurIPS},
  volume={33},
  pages={1877--1901},
  year={2020}
}

@article{FLAN,
  title={Finetuned language models are zero-shot learners},
  author={Wei, Jason and Bosma, Maarten and Zhao, Vincent Y and Guu, Kelvin and Yu, Adams Wei and Lester, Brian and Du, Nan and Dai, Andrew M and Le, Quoc V},
  journal={arXiv preprint arXiv:2109.01652},
  year={2021}
}

@article{Tzero,
  title={Multitask prompted training enables zero-shot task generalization},
  author={Sanh, Victor and Webson, Albert and Raffel, Colin and Bach, Stephen H and Sutawika, Lintang and Alyafeai, Zaid and Chaffin, Antoine and Stiegler, Arnaud and Scao, Teven Le and Raja, Arun and others},
  journal={arXiv preprint arXiv:2110.08207},
  year={2021}
}

@article{Llama,
  title={Llama: Open and efficient foundation language models},
  author={Touvron, Hugo and Lavril, Thibaut and Izacard, Gautier and Martinet, Xavier and Lachaux, Marie-Anne and Lacroix, Timoth{\'e}e and Rozi{\`e}re, Baptiste and Goyal, Naman and Hambro, Eric and Azhar, Faisal and others},
  journal={arXiv preprint arXiv:2302.13971},
  year={2023}
}

@inproceedings{synthetic_apple_VA,
  title={Synthetic query generation using large language models for virtual assistants},
  author={Sannigrahi, Sonal and Fraga-Silva, Thiago and Oualil, Youssef and Van Gysel, Christophe},
  booktitle={SIGIR},
  pages={2837--2841},
  year={2024}
}

@inproceedings{lo2020s2orc,
  title={S2ORC: The Semantic Scholar Open Research Corpus},
  author={Lo, Kyle and Wang, Lucy Lu and Neumann, Mark and Kinney, Rodney and Weld, Daniel S},
  booktitle={ACL},
  pages={4969--4983},
  year={2020}
}

@article{nogueira2019doc2query,
  title={From doc2query to docTTTTTquery},
  author={Nogueira, Rodrigo and Lin, Jimmy},
  journal={Online preprint},
  volume={6},
  number={2},
  year={2019}
}

@inproceedings{DensePRF,
  title={Improving query representations for dense retrieval with pseudo relevance feedback},
  author={Yu, HongChien and Xiong, Chenyan and Callan, Jamie},
  booktitle={CIKM},
  pages={3592--3596},
  year={2021}
}

@inproceedings{razdaibiedina2023miread,
  title={MIReAD: Simple Method for Learning High-quality Representations from Scientific Documents},
  author={Razdaibiedina, Anastasiia and Brechalov, Aleksandr},
  booktitle={ACL},
  pages={530--539},
  year={2023}
}

@article{liang2020embedding,
  title={Embedding-based zero-shot retrieval through query generation},
  author={Liang, Davis and Xu, Peng and Shakeri, Siamak and Santos, Cicero Nogueira dos and Nallapati, Ramesh and Huang, Zhiheng and Xiang, Bing},
  journal={arXiv preprint arXiv:2009.10270},
  year={2020}
}

@inproceedings{ma2021zero,
  title={Zero-shot Neural Passage Retrieval via Domain-targeted Synthetic Question Generation},
  author={Ma, Ji and Korotkov, Ivan and Yang, Yinfei and Hall, Keith and McDonald, Ryan},
  booktitle={EACL},
  pages={1075--1088},
  year={2021}
}

@inproceedings{wang2022gpl,
  title={GPL: Generative Pseudo Labeling for Unsupervised Domain Adaptation of Dense Retrieval},
  author={Wang, Kexin and Thakur, Nandan and Reimers, Nils and Gurevych, Iryna},
  booktitle={NAACL},
  pages={2345--2360},
  year={2022}
}

@inproceedings{taxoindex,
  title={Taxonomy-guided Semantic Indexing for Academic Paper Search},
  author={Kang, SeongKu and Zhang, Yunyi and Jiang, Pengcheng and Lee, Dongha and Han, Jiawei and Yu, Hwanjo},
    booktitle = {EMNLP},
    pages = {7169--7184},
  year={2024}
}

@inproceedings{hyde,
  title={Precise zero-shot dense retrieval without relevance labels},
  author={Gao, Luyu and Ma, Xueguang and Lin, Jimmy and Callan, Jamie},
  booktitle={ACL},
  pages={1762--1777},
  year={2023}
}

@inproceedings{ccqgen,
  title={Improving scientific document retrieval with concept coverage-based query set generation},
  author={Kang, SeongKu and Jin, Bowen and Kweon, Wonbin and Zhang, Yu and Lee, Dongha and Han, Jiawei and Yu, Hwanjo},
  booktitle={WSDM},
  pages={895--904},
  year={2025}
}

@article{doc2query,
  title={Document expansion by query prediction},
  author={Nogueira, Rodrigo and Yang, Wei and Lin, Jimmy and Cho, Kyunghyun},
  journal={arXiv preprint arXiv:1904.08375},
  year={2019}
}

@inproceedings{query2doc,
  title={Query2doc: Query Expansion with Large Language Models},
  author={Wang, Liang and Yang, Nan and Wei, Furu},
  booktitle={EMNLP},
  pages={9414--9423},
  year={2023}
}

@inproceedings{doc2querymm,
  title={Revisiting document expansion and filtering for effective first-stage retrieval},
  author={Mansour, Watheq and Zhuang, Shengyao and Zuccon, Guido and Mackenzie, Joel},
  booktitle={SIGIR},
  pages={186--196},
  year={2024}
}

@article{chen2018snippet,
  title={Modeling queries with contextual snippets for information retrieval},
  author={Chen, Qin and Hu, Qinmin and Huang, Jimmy Xiangji and He, Liang},
  journal={ACM Transactions on Intelligent Systems and Technology},
  year={2018},
  volume={9},
  number={4},
  pages={1--26},
   
}

@article{ContrastiveSurvey2023,
  author = {Xu, Lingling and Xie, Haoran and Li, Zongxi and Wang, Fu Lee and Wang, Weiming and Li, Qing},
  title = {Contrastive Learning Models for Sentence Representations: A Survey},
  journal = {ACM Transactions on Intelligent Systems and Technology},
  year = {2023},
  volume = {14},
  number = {4},
  pages = {1--34}
}

@inproceedings{condenser,
  title={Condenser: a Pre-training Architecture for Dense Retrieval},
  author={Gao, Luyu and Callan, Jamie},
  booktitle={EMNLP},
  year={2021}
}

@article{corank,
  title={CoRank: LLM-based compact reranking with document features for scientific retrieval},
  author={Tian, Runchu and Xu, Xueqiang and Jin, Bowen and Kang, SeongKu and Han, Jiawei},
  journal={arXiv preprint arXiv:2505.13757},
  year={2025}
}

@inproceedings{kweon2026pairsem,
  title={PairSem: LLM-Guided Pairwise Semantic Matching for Scientific Document Retrieval},
  author={Kweon, Wonbin and Tian, Runchu and Kang, SeongKu and Jiang, Pengcheng and Lu, Zhiyong and Han, Jiawei and Yu, Hwanjo},
  booktitle={WWW},
  pages={2396--2407},
  year={2026}
}

@inproceedings{semrank,
  title={Scientific paper retrieval with llm-guided semantic-based ranking},
  author={Zhang, Yunyi and Yang, Ruozhen and Jiao, Siqi and Kang, SeongKu and Han, Jiawei},
  booktitle={Findings of EMNLP},
  pages={2049--2060},
  year={2025}
}

@inproceedings{zhang2025chain,
  title={Chain-of-factors paper-reviewer matching},
  author={Zhang, Yu and Shen, Yanzhen and Kang, SeongKu and Chen, Xiusi and Jin, Bowen and Han, Jiawei},
  booktitle={WWW},
  pages={1901--1910},
  year={2025}
}
